\documentstyle[11pt]{article}

\font\msbm=msbm10 at 11pt
\font\eufm=eufm10 at 11pt
\font\eusm=eusm10 at 11pt
\font\eufma=eufm8 at 8pt
\font\msbma=msbm8 at 8pt
\font\eufms=eufm6 at 6pt
\font\msbms=msbm6 at 6pt

\newtheorem{definition}{Definition}
\newtheorem{proposition}{Proposition}

\newtheorem{lemma}{Lemma}

\newcommand{\CC}{\mbox{{\msbm C}}}
\newcommand{\RR}{\mbox{{\msbm R}}}

\newcommand{\NN}{\mbox{{\msbm N}}}
\newcommand{\ZZg}{\mbox{{\msbm Z}$_{2}$}}
\newcommand{\NNN}{\mbox{{\msbm N}$_{0}$}}
\newcommand{\CCa}{\mbox{{\msbma C}}}

\newcommand{\NNa}{\mbox{{\msbma N}}}

\newcommand{\NNNa}{\mbox{{\msbma N}}_{0}}

\newcommand{\NNs}{\mbox{{\msbms N}}}

\newcommand{\NNNs}{\mbox{{\msbms N}}_{0}}
\newcommand{\CCn}[1]{\mbox{{\msbm C}$^{#1}$}}
\newcommand{\RRn}[1]{\mbox{{\msbm R}$^{#1}$}}

\newcommand{\SUB}{\mbox{{\msbm S}}}
\newcommand{\SUBa}{\mbox{{\msbma S}}}

\newcommand{\DS}{\displaystyle}
\newcommand{\BD}{\begin{displaymath}}
\newcommand{\ED}{\end{displaymath}}
\newcommand{\BE}{\begin{enumerate}}
\newcommand{\EE}{\end{enumerate}}
\newcommand{\BA}{\begin{eqnarray*}}
\newcommand{\EA}{\end{eqnarray*}}
\newcommand{\BG}{\begin{equation}}
\newcommand{\EG}{\end{equation}}
\newcommand{\BGG}{\begin{eqnarray}}
\newcommand{\EGG}{\end{eqnarray}}

\newcommand{\lp}{\left (}
\newcommand{\rp}{\right )}
\newcommand{\lb}{\left [}
\newcommand{\rb}{\right ]}
\newcommand{\lbr}{\left \{}
\newcommand{\rbr}{\right \}}

\newcommand{\norm}[1]{\left\vert\!\left\vert #1\right\vert\!\right\vert}
\newcommand{\semn}[3]{\left\vert#1\right\vert^{#2}_{#3}}
\newcommand{\Comu}[2]{\left [#1,#2\right ]}
\newcommand{\Comug}[2]{\left [#1,#2\right ]_{g}}
\newcommand{\scalar}[2]{\left\langle #1\left\vert #2\right.\right\rangle}
\newcommand{\scalarl}[2]{\left\langle\left. #1\right\vert #2\right\rangle}
\newcommand{\scalarK}[3]{\left\langle #1\left\vert #2\right.\right\rangle_{#3}}

\newcommand{\hwedge}{\hat{\wedge}}

\newcommand{\Over}[1]{\overline{#1}}
\newcommand{\rom}[1]{{\rm #1}}
\newcommand{\Quot}[1]{\mbox{\boldmath $#1$\unboldmath}}
\newcommand{\Quotgr}[1]{\mbox{\boldmath $#1$\unboldmath}}
\newcommand{\Quota}[1]{\mbox{\footnotesize\boldmath $#1$\unboldmath}}

\newcommand{\Oger}[1]{\mbox{{\eufm #1}}}

\newcommand{\Indn}[1]{\mbox{{\eufm I}$_{#1}$}}
\newcommand{\GIndn}[1]{\mbox{{\eufm I}$_{#1}^{(3\vert 2)}$}}
\newcommand{\Indna}[1]{\mbox{{\eufma I}}_{#1}}
\newcommand{\GIndna}[1]{\mbox{{\eufm I}}_{#1}^{(3\vert 2)}}
\newcommand{\Indns}[1]{\mbox{{\eufms I}}_{#1}}
\newcommand{\OIndn}[1]{\mbox{{\eufm I}$_{#1}^{(3)}$}}
\newcommand{\OIndna}[1]{\mbox{{\eufm I}}_{#1}^{(3)}}

\newcommand{\Vecs}[2]{B^{#1\vert #2}_{L}}
\newcommand{\Vecsg}[3]{B^{#1\vert #2}_{L,#3}}

\newcommand{\Ss}{S_{\tiny\rho}}
\newcommand{\NSs}[1]{{\cal S}_{\tiny\rho,#1}}
\newcommand{\Sp}{{\rm S}_{\tiny\rho}}
\newcommand{\NSp}[1]{\mbox{{\eusm S}}_{\tiny\rho,#1}}
\newcommand{\Idss}{{\cal I}_{\Ss}}
\newcommand{\Idsp}{{\cal I}_{\Sp}}

\newcommand{\Hu}{${\cal H}^{\infty}$}
\newcommand{\Huor}[1]{{\cal H}^{\infty}(#1,B_{L})}
\newcommand{\Huoc}[1]{{\cal H}^{\infty}(#1,C_{L})}

\newcommand{\Cu}{${\cal C}^{\infty}$}

\newcommand{\Cuoc}[1]{{\cal C}^{\infty}(#1,\CC)}

\newcommand{\Poloc}[1]{{\cal P}(#1)}

\newcommand{\Polocn}[2]{{\cal P}^{#1}(#2)}

\newcommand{\Ort}[2]{\mbox{{\eufm osp}}(#1\vert #2)}
\newcommand{\Ortg}[3]{\mbox{{\eufm osp}}(#1\vert #2)_{\Over{#3}}}

\newcommand{\Plv}[1]{\mbox{{\eufm pl}}(#1)}

\newcommand{\Sl}[1]{\mbox{{\eufm sl}}(#1)}

\newcommand{\Perm}[1]{\mbox{{\eufm S}}_{#1}}
\newcommand{\Perma}[1]{\mbox{{\eufma S}}_{#1}}

\newcommand{\Svec}[1]{\Oger{V}^{g}(#1)}
\newcommand{\Vec}[1]{\Oger{V}(#1)}
\newcommand{\Svecg}[2]{\Oger{V}^{g}(#1)_{\Over{#2}}}
\newcommand{\Derc}[1]{\Oger{Der}_{\CCa}^{g}(#1)}

\newcommand{\DER}[1]{\Oger{Der}_{\CCa}(#1)}

\newcommand{\Sphar}[4]{\Quotgr{Y}^{#1}_{#2,#3,\Over{#4}}}
\newcommand{\Sphard}[5]{\Quotgr{Y}^{#1}_{#2,#3,\Over{#4}+\Over{#5}}}
\newcommand{\VSphar}[1]{\Quotgr{V}^{#1}}
\newcommand{\NSphar}[5]{Y^{(#1)\, #2}_{#3,#4,\Over{#5}}}
\newcommand{\NSphard}[6]{Y^{(#1)\, #2}_{#3,#4,\Over{#5}+\Over{#6}}}
\newcommand{\NVSphar}[2]{V^{(#1)\, #2}}
\newcommand{\Sh}[2]{{\bf Y}^{#1}_{#2}}
\newcommand{\NSh}[3]{\rom{Y}^{(#1)\, #2}_{#3}}

\newcommand{\End}[1]{{\rm End}_{\CCa}(#1)}

\newcommand{\Cent}[1]{{\cal Z}^{g}(#1)}
\newcommand{\Centc}[1]{{\cal Z}(#1)}
\newcommand{\Centg}[2]{{\cal Z}^{g}(#1)_{\Over{#2}}}
\newcommand{\Hrpmm}[4]{{\rm Hom}_{#1}^{#2}(#3;#4)}

\newcommand{\SDC}[1]{\Omega^{g}({\cal S}_{\tiny\rho,#1})}
\newcommand{\SDCp}[2]{\Omega^{g,#1}({\cal S}_{\tiny\rho,#2})}

\newcommand{\SDCpg}[3]{\Omega^{g,#1}({\cal S}_{\tiny\rho,#2})_{\Over{#3}}}
\newcommand{\SDCc}[1]{\Omega^{g}_{{\cal Z}^{g}}({\cal S}_{\tiny\rho,#1})}
\newcommand{\SDCpc}[2]{\Omega^{g,#1}_{{\cal Z}^{g}}({\cal S}_{\tiny\rho,#2})}

\newcommand{\DC}[1]{\Omega(\NSp{#1})}
\newcommand{\DCp}[2]{\Omega^{#1}(\NSp{#2})}
\newcommand{\DCc}[1]{\Omega_{{\cal Z}}(\NSp{#1})}
\newcommand{\DCpc}[2]{\Omega^{#1}_{{\cal Z}}(\NSp{#2})}

\begin{document}

\pagenumbering{arabic}
\vspace*{0.3cm}

$\qquad \qquad \qquad \qquad \qquad \qquad \qquad \qquad \qquad \qquad \qquad \qquad \qquad \qquad \qquad \qquad$ math-ph/9804013

$\qquad \qquad \qquad \qquad \qquad \qquad \qquad \qquad \qquad \qquad \qquad \qquad \qquad \qquad \qquad \qquad$ mp-arc/98-283

$\qquad \qquad \qquad \qquad \qquad \qquad \qquad \qquad \qquad \qquad \qquad \qquad \qquad \qquad \qquad \qquad$ ESI 546

\vspace{0.9cm}

\LARGE
\centerline{{\bf The Fuzzy Supersphere}}

\vspace{1cm}

\Large
\centerline{H.Grosse$^{\rom{a},}$\footnote{E-mail: grosse@doppler.thp.univie.ac.at}
and G.Reiter$^{\rom{b},}$\footnote{Research partly supported by the
``Fonds zur F\"orderung der wissenschaftlichen Forschung (FWF)''
through the research project P11783-PHY ``Quantum Field Theory on Noncommutative Manifolds''.}$^{,}$\footnote{Corresponding author: Tel.: +43 316 873 8670, Fax:
+43 316 873 8678, E-mail: reiter@itp.tu-graz.ac.at}}

\vspace{7mm}

\normalsize
\centerline{\parbox{115mm}
{$^{\rom{a}}$ Universit\"{a}t Wien, Institut f\"ur Theoretische Physik, Boltzmanngasse 5,
A-1090 Wien, Austria}}

\vspace{2mm}

\centerline{\parbox{115mm}
{$^{\rom{b}}$ Technische Universit\"{a}t Graz, Institut f\"ur Theoretische Physik, Petersgasse 16, A-8010 Graz, Austria}}

\vspace{1cm}

\large
\centerline{{\bf Abstract}}
\normalsize

\vspace{4mm}

\centerline{\parbox{140mm}
{We introduce the fuzzy supersphere as sequence of finite-dimensional,
noncommutative \ZZg-graded algebras tending in a suitable limit to a dense
subalgebra of the \ZZg-graded algebra of \Hu-functions on the
$(2\vert 2)$-dimensional supersphere. Noncommutative analogues of the
body map (to the (fuzzy) sphere) and the super-deRham complex are introduced.
In particular we reproduce the equality of the super-deRham cohomology of the supersphere and the ordinary deRham cohomology of its body on the ``fuzzy level''.
\begin{tabbing}
\> geometry, fuzzy sphere
\end{tabbing}}}

\section{Introduction}

Thinking about space and time noncommutative geometry \cite{Connes1} offers an enormous
general framework for physical model building, because one can get rid of the concept
of points. The basic idea of noncommutative geometry is to formulate first notions
on differentiable manifolds in terms of their commutative \CC-algebras of differentiable,
complex-valued functions in order to generalize subsequently these notions, which do not
depend on the commutativity to abstract, not necessarily commutative algebras. So in general
one will lose the notion of points (corresponding in ordinary geometry to the spectrum of the commutative algebra of functions) and the role of general, noncommutative manifolds
is played by abstract algebras.\newline
Of course there is no canonical way how to associate a noncommutative algebra with some
mathematical or physical model of spacetime, phase space or some more ``exotic'' objects and one can contrive lots of different procedures. Beside that of fuzzy manifolds \cite{Madore2,Grosse7,Grosse13,Hawkins1}, which is intimately related to quantization and which we will follow in the sequel, let us also mention (without being complete) a similar approach for the Minkowski space \cite{Doplicher1},
quantum group motivated approaches (see for example \cite{Manin1}) and approaches based on
posets (see for example \cite{Balachandran1}). Fuzzy manifolds are not only \CC-algebras
but whole sequences (or more precisely directed systems) of noncommutative \CC-algebras,
which approximate in a very specific way the corresponding ordinary manifolds. Each of the
\CC-algebras of such a sequence can be interpreted as description of the corresponding
ordinary manifold on which localization is possible only up to a minimal length.
By employing the tools of noncommutative geometry and of
matrix geometry \cite{Dubois-Violette1} in particular, it was possible to introduce on a
specific fuzzy manifold, namely the fuzzy sphere, (sections of) vector bundles, a differential calculus, an integral, etc. and to use this for the formulation of field theoretical
models \cite{Madore2,Madore4,Grosse1,Grosse5,Grosse14,Klimcik1}. When these models are quantized a very interesting feature shows up: They are finite and the fuzziness plays the role of a non-perturbative regulator, which does not break the characteristic symmetries of the corresponding continuum theories.\newline
Although supermanifolds are to some extend ``baby-noncommutative geometries'' they are
treated and interpreted in the spirit of classical differential and algebraic geometry.
Localization is (depending on the approach to supermanifolds) more or less present
and the term ``super'' should rather be seen as additional structure. Noncommutative
generalizations should be described by \ZZg-graded algebras over a ring, that depends
on the class of supermanifolds under consideration. We want to mention, that
there exist already several articles and books in the literature dealing with various aspects of \ZZg-graded \CC-algebras, supersymmetry and noncommutative geometry. Without being
complete let us just mention \cite{Kastler1}, where notions as cyclic cohomology and Fredholm
modules are treated in the \ZZg-graded setting, \cite{Dubois-Violette2,Kerner1}, where a possibility of generalizing matrix geometry to the \ZZg-graded framework is presented and \cite{Kalau1}, where the concept of a spectral triple is extended to algebras which contain bosonic and fermionic degrees of freedom.\newline
Our aim in this article is to develop first a ``fuzzy variant'' of the $(2\vert 2)$-dimensional supersphere and subsequently an analogue to the super-deRham complex on each of the resulting \ZZg-graded \CC-algebras. Motivated by the wish to find an adequate
language for the description of the spinor bundle on the fuzzy sphere one of the authors
together with C.Klim\v{c}ik and P.Pre\v{s}najder already solved the first 
problem \cite{Grosse1}. Here we want to embed the description given there a little bit
more in the language of supermanifolds in the sense of A.Rogers \cite{Rogers1,Rogers7} as well as that of graded manifolds \cite{Kostant1,Leites1} in order to have a good guideline for the development of a \ZZg-graded differential calculus later on.

More precisely the article is organized as follows. In the first two chapters we describe
the $(2\vert 2)$-dimensional supersphere as \Hu-deWitt super- respectively graded manifold
and establish on its \ZZg-graded algebra of complex-Grassmann-valued functions additional
structures such as a Fr\'{e}chet topology, an indefinite scalar product and a ``completely reducible'' grade star representation of the orthosymplectic Lie superalgebra $\Ort{1}{2}$. The article is written as self-contained as possible, but for the basics of the theory of supermanifolds we refer the reader to the above cited original articles as well as to the excellent book \cite{Bartocci1}.\newline
In chapter 4 we endow each element of a specific sequence of submodules of the 
$\Ort{1}{2}$-module of complex-Grassmann-valued \Hu-functions on the $(2\vert 2)$-dimensional supersphere with a new \ZZg-graded product and define by this the fuzzy supersphere. The graded-commutative limit of these products is proven. 
The body of the $(2\vert 2)$-dimensional supersphere is the
ordinary $2$-sphere and chapter 5 is devoted to extend the corresponding body map 
to the fuzzy setting.\newline
In chapter 6 we define on each of the ``truncated superspheres'' (in the sense explained
above) an analogue to the super-deRham complex by transposing the idea of derivation-based differential calculi \cite{Dubois-Violette1,Dubois-Violette6} and its extension used for the definition of a differential calculus on the fuzzy sphere \cite{Madore2,Madore10} to the 
\ZZg-graded setting. The resulting complex is nothing else but the complex of Lie superalgebras with coefficients in a non-trivial $\Ort{1}{2}$-module and it is 
infinite-dimensional, as it is usual for supermanifolds. The latter fact shows in particular,
that the complex is completely different to that proposed in 
\cite{Dubois-Violette2,Kerner1}. Our construction is natural in the sense, that the noncommutative body map
extends - as it is the case for graded manifolds - to a cochain map from the algebra of superforms on the truncated supersphere to the algebra of forms on its body.\newline 
The final chapter is devoted to the calculation of the cohomology corresponding
with the differential complex. In particular we reproduce the equality of the super-deRham cohomology of the supersphere and the ordinary deRham cohomology of its body on the ``fuzzy level''.

Let us finally mention some conventions which are used throughout the article.
When we speak of algebras we always mean associative algebras with identity; algebra homomorphisms always preserve the identity. Left or right modules over an algebra are always assumed to be unital.\newline
There will appear lots of \ZZg-graded objects. If the object is denoted by ${\cal O}$ its
even part is denoted by ${\cal O}_{\Over{0}}$, its odd part by ${\cal O}_{\Over{1}}$.
If $e$ is some homogeneous element of such an object its degree will be denoted by $\Over{e}$. 
Speaking of grading in the context of an ungraded object we mean, that the object is endowed
with its trivial graduation. Moreover a left or right \ZZg-graded module over a \ZZg-graded,
graded-commutative algebra is always assumed to be given its canonical \ZZg-graded
bimodule structure.\newline
There is a one-to-one correspondence between representations of a Lie (super)algebra and its
universal enveloping algebra: We do not distinguish between these representations and
pass freely from the language of representations to the language of modules and 
vice versa. 

\section{The (2$\vert$2)-dimensional supersphere}

In this preliminary section we introduce the (2$\vert$2)-dimensional supersphere as~\Hu-deWitt supermanifold (and by this also as graded manifold) and characterize the \ZZg-graded algebra of~\Hu-functions on it as a suitable quotient of the algebra of~\Hu-functions on the (3$\vert$2)-dimensional vector superspace.
Moreover we endow this algebra with additional structures, such as a Fr\'{e}chet topology  
and an indefinite scalar product, which we will need later on for the formulation and proof
of the ``graded-commutative limit'' of the fuzzy supersphere. 

For $L\in\NN$~let $B_{L}$ denote the Grassmann algebra over $\RRn{L}$ and $C_{L}$ the Grassmann algebra over $\CCn{L}$. We view both of them as \ZZg-graded algebras. Furthermore we introduce the set
\BG
\label{s1}
\Indn{L} := \lbr M=(i_{1},\cdots,i_{p})\vert~i_{1},\cdots,i_{p}=
1,\cdots,L;~1\leq p\leq L~\mbox{with}~i_{1}<\cdots<i_{p}\rbr
\cup\lbr\emptyset\rbr \; .
\EG
If $\{e_{i}\}_{i=1,\cdots,L}$ is a basis of $\RRn{L}$ (or $\CCn{L}$) then a homogeneous basis of $B_{L}$ (or $C_{L}$) is formed by 
the elements
\BGG
\label{s2}
e_{M} &:=& e_{i_{1}}e_{i_{2}}\cdots e_{i_{p}}
\quad \mbox{with} \quad M = (i_{1},i_{2},\cdots,i_{p}) \in \Indn{L}
\nonumber \\
e_{\emptyset} &:=& 1 
\EGG
and
\BG
\label{s3}
\norm{y} \equiv \norm{\sum_{M\in\Indna{L}}y_{M}e_{M}} :=
\sum_{M\in\Indna{L}}\left\vert y_{M}\right\vert 
\EG
for every $y\in B_{L}(C_{L})$, defines a norm on $B_{L}$ (or $C_{L}$). The \RR-linear map
$B_{L}\longrightarrow C_{L}$, defined by identifying the basis elements (\ref{s2}) of $B_{L}$ with the corresponding basis elements of $C_{L}$, is an injective isometric homomorphism
of \ZZg-graded \RR-Banach algebras, which we will use to interpret $B_{L}$ as closed and graded \RR-subalgebra of $C_{L}$. Consequently the elements of $B_{L}$ are those elements of $C_{L}$, which are invariant with respect to complex conjugation 
$*:C_{L}\longrightarrow C_{L}$, defined by
\BG
\label{s3a}
y^{*} \equiv \lp\sum_{M\in\Indna{n}}y_{M}e_{M}\rp^{*} :=
\sum_{M\in\Indna{n}}y^{*}_{M}e_{M} \, .
\EG
The direct sum of $B_{L,\Over{0}}$-modules
\BG
\label{s4}
\Vecs{r}{s} :=  B_{L,\Over{0}}\oplus\stackrel{r}{\cdots}
\oplus B_{L,\Over{0}}\oplus B_{L,\Over{1}}\oplus\stackrel{s}{\cdots}
\oplus B_{L,\Over{1}} \, , \qquad r,s\in\NNN \, ,
\EG
together with the \ZZg-grading
\BGG
\label{s5}
\Vecsg{r}{s}{\Over{0}} := B_{L,\Over{0}}\oplus\stackrel{r}{\cdots}
\oplus B_{L,\Over{0}} \equiv \Vecs{r}{0}
\nonumber \\
\Vecsg{r}{s}{\Over{1}} := B_{L,\Over{1}}\oplus\stackrel{s}{\cdots}
\oplus B_{L,\Over{1}} \equiv \Vecs{0}{s}
\EGG
and the norm
\BG 
\label{s6}
\norm{(x^{k},\theta^{\alpha})} :=
\sum_{k=1}^{r}\norm{x^{k}} + \sum_{\alpha=r+1}^{r+s}\norm{\theta^{\alpha}} \, ,
\qquad (x^{k},\theta^{\alpha})\in\Vecs{r}{s} \, ,
\EG
is called $(r\vert s)$-dimensional real vector superspace.\newline
The body (or augmentation) map on $C_{L}$ will be denoted by $\epsilon$;
the body map $\Vecs{r}{s}\longrightarrow\RRn{r}$, $(x^{k},\theta^{\alpha})\mapsto
(\epsilon(x^{k}))$ by $\Phi_{r\vert s}$. Beside the topology induced
by the norm (\ref{s6}), the so-called fine topology, there is another important topology on the vector superspace: the coarse or deWitt-topology. By definition it is the coarsest topology on $\Vecs{r}{s}$ such that the body map $\Phi_{r\vert s}$ is continuous.

For an arbitrary $\rho\in\RR^{+}$ we define the $(2\vert 2)$-dimensional supersphere $\Ss$ of ``radius'' $\rho$ as the closed topological subspace (with respect to the fine topology) of all points of $\Vecs{3}{2}$~($L\geq 2$, fixed)~fulfilling 
\BG 
\label{s7}
P_{\rho}(x^{k},\theta^{\alpha}) :=
\sum_{i=1}^{3}(x^{i})^{2} + 2\theta^{4}\theta^{5} - \rho^{2} = 0 \, .
\EG
$\Ss$ can be endowed with the structure of a $(2\vert 2)$-dimensional
\Hu-deWitt supermanifold (see also \cite{Bartocci10,Bryant5,Crane1,Teofilatto1}). In order to do so one introduces the ``north'' and
``south pole-$\epsilon$-fibers''
\BG
\label{s8}
F_{\pm} := \lbr (x^{k},\theta^{\alpha})\in\Ss 
\vert~\Phi_{3\vert 2}(x^{k},\theta^{\alpha})=(0,0,\pm\rho) \rbr
\EG
as well as their open complements 
\BG
\label{s9}
U_{\pm} := \Ss\backslash F_{\pm}
\EG
in $\Ss$. Then the superstereographic projections
\BGG
\label{s10}
h_{\pm}: U_{\pm} \longrightarrow \Vecs{2}{2}
\qquad \qquad \qquad \qquad \quad
\\
(x^{k},\theta^{\alpha}) \mapsto
\frac{\rho}{\rho\mp x^{3}}(x^{1},x^{2},\theta^{4},\theta^{5})
\nonumber 
\EGG
form a~\Hu-deWitt atlas on $\Ss$, because the transition functions
\BG
\label{s11}
h_{\pm}\circ h^{-1}_{\mp}(y^{j},\eta^{\alpha}) =
\frac{\rho^{2}}{(y^{1})^{2}+(y^{2})^{2}+2\eta^{3}\eta^{4}}
(y^{j},\eta^{\alpha})
\EG
are~\Hu-functions $\Vecs{2}{2}\backslash\Phi_{2\vert 2}^{-1}(0,0)
\longrightarrow\Vecs{2}{2}\backslash\Phi_{2\vert 2}^{-1}(0,0)$ on the one hand and
$h_{\pm}(U_{\pm})$ are coarse open on the other hand. The corresponding (\Cu-)body manifold
can be identified canonically with the $2$-dimensional sphere $\Sp$ of radius $\rho$ embedded in $\RRn{3}$ and the body projection $\Phi_{S}$ is then simply given by
\BG
\label{s12}
\Phi_{S} = \Phi_{3\vert 2}\Big\vert_{\Ss} \, .
\EG
Let us denote the sheaf of real- and complex Grassmann-valued~\Hu-functions on a 
\Hu-supermanifold $X$ by $\Huor{-}$ and $\Huoc{-}$, respectively. We view $\Huor{-}$ as 
subsheaf of $\Huoc{-}$ in the natural way. That is, for every open subset $U$ of the
super\-ma\-ni\-fold $X$ we define complex conjugation 
$*:\Huoc{U}\longrightarrow\Huoc{U}$ 
pointwise and characterize $\Huor{U}$ as the graded \RR-subalgebra of $*$-invariant elements.
It should be noted, that, if $X$ is a \Hu-deWitt supermanifold with body $\rom{X}$ and
body projection $\Phi_{X}$, the pair $(\rom{X},\Phi_{X*}\Huor{-})$, where $\Phi_{X*}\Huor{-}$ denotes the direct image of $\Huor{-}$ by $\Phi_{X}$, is a graded manifold \cite{Bartocci1}. Consequently the map $\beta_{X}$ from the \ZZg-graded \CC-algebra $\Huoc{X}$ to the (trivially 
\ZZg-graded) \CC-algebra $\Cuoc{\rom{X}}$ of smooth complex-valued functions on $\rom{X}$, defined by 
\BG
\label{s13}
\lp\beta_{X}(f)\rp(\Phi_{X}(x)) := \epsilon\circ f(x)
\EG
for all $x\in X$, is a surjective homomorphism of graded algebras.

The set
\BG
\label{s14}
\Idss := \lbr~f\in\Huoc{\Vecs{3}{2}}~\left\vert~f\Big\vert_{\Ss}=0\right.\rbr
\EG
of all complex-Grassmann-valued functions on $\Vecs{3}{2}$
vanishing on the supersphere is a graded ideal in $\Huoc{\Vecs{3}{2}}$. 
We will identify $\Huoc{\Ss}$ and $\Huoc{\Vecs{3}{2}}/\Idss$ according to the first part of the following result.
\begin{lemma} 
\label{lemma1s}
The map $\chi:\Huoc{\Vecs{3}{2}}/\Idss\longrightarrow\Huoc{\Ss}$, defined
by
\BG
\label{s15}
\chi(\Quot{f}) := f\Big\vert_{\Ss} \, , \quad f\in\Quot{f}\, ,
\EG
is an isomorphism of \ZZg-graded \CC-algebras. Moreover, for every $f\in\Idss$ there 
exists a \Hu-function 
$g\in\Huoc{\Vecs{3}{2}}$, such that
\BG
\label{s15a}
f = P_{\rho}g
\EG
is fulfilled.
\end{lemma}
${\sl Proof}$: Analogous to (\ref{s8}),(\ref{s9}) one introduces
``north'' and ``south pole-$\epsilon$-fibers'' 
$\tilde{F}_{\pm}:=\Phi^{-1}_{3\vert 2}(0,0,\RR^{\pm}_{0})$ as well as
their open complements $\tilde{U}_{\pm}:=\Vecs{3}{2}\setminus\tilde{F}_{\pm}$ in $\Vecs{3}{2}$. Then the maps 
\BGG
\label{s16}
\tilde{h}_{\pm}&:& \tilde{U}_{\pm} \longrightarrow 
\Phi_{3\vert 2}^{-1}(\RRn{2}\times\RR^{+})
\nonumber
\\
&& (x^{k},\theta^{\alpha}) \mapsto
\frac{\sqrt{\sum_{i=1}^{3}(x^{i})^{2}+2\theta^{4}\theta^{5}}}
{\sqrt{\sum_{i=1}^{3}(x^{i})^{2}+2\theta^{4}\theta^{5}}\mp x^{3}}\cdot
\\
&& \qquad \qquad \quad \cdot
(x^{1},x^{2},\lp\sqrt{\sum_{i=1}^{3}(x^{i})^{2}+2\theta^{4}\theta^{5}}\mp
x^{3}\rp\sqrt{\sum_{i=1}^{3}(x^{i})^{2}+2\theta^{4}\theta^{5}},\theta^{4},\theta^{5}) \, ,
\nonumber
\EGG
where the square root is defined via the ${\rm Z}$-expansion of the
ordinary square root, are ``subsupermanifold charts'' of the
vector superspace, which one can use to conclude $f\vert_{\Ss}\in\Huoc{\Ss}$.
Obviously $\chi$ is an injective, even homomorphism of \ZZg-graded algebras and the
surjectivity of $\chi$ follows from the existence of \Hu-partitions of the unity on
coarse open coverings of $\Vecs{3}{2}$ \cite{Bartocci1}.\newline
Using once again such a coarse \Hu-partitions of the unity one can conclude, that it is 
enough to show (\ref{s15a}) on $\tilde{U}_{\pm}$. The next step is to note, that
the condition $f\vert_{\tilde{U}_{\pm}\cap\Ss}=0$ means
$f\circ\tilde{h}_{\pm}^{-1}\vert_{y^{3}=\rho^{2}} = 0$ 
for the functions 
$f\circ\tilde{h}_{\pm}^{-1}\in\Huoc{\Phi_{3\vert 2}^{-1}(\RRn{2}\times\RR^{+})}$ and that it is enough to prove
\BD
f\circ\tilde{h}_{\pm}^{-1} = \lp y^{3}-\rho^{2}\rp g_{\pm} \, , \quad
g_{\pm}\in\Huoc{\Phi_{3\vert 2}^{-1}(\RRn{2}\times\RR^{+})} \, .
\ED
But if one exploits the properties of the ``superfield'' and Z-expansion one reduces the
above problem to the same problem on the ``\Cu-level'', which can be solved in the usual 
way (see \cite{Bartocci1,Constantinescu1}) via integration.
\hfill $\Box$

\vspace{2mm}

Analogous to the case of the supersphere we can identify the \CC-algebra $\Cuoc{\Sp}$ with
the \CC-algebra $\Cuoc{\RRn{3}}$ factored by the ideal $\Idsp$ of all \Cu-functions on
$\RRn{3}$ vanishing on the ordinary 2-sphere $\Sp$. Because of (\ref{s12}) and
\BG
\label{s16a}
\beta_{3\vert 2}\lp\Idss\rp \subseteq \Idsp \, ,
\EG
the ``algebraic body map'' $\beta_{\Ss}:\Huoc{\Ss}\longrightarrow\Cuoc{\Sp}$ is simply determined by
\BG
\label{s16b}
\beta_{\Ss}(\Quotgr{f}) = {\bf f} \, ,
\EG
where we introduced the notation ${\bf f}$ for the equivalence class of $\beta_{3\vert 2}(f)$
for some $f\in\Quotgr{f}$.

In the spirit of the theory of graded manifolds we introduce the set $\Svec{X}$ of complex global supervector fields on a \Hu-deWitt supermanifold $X$ as the set of graded
derivations $\Derc{\Huoc{X}}$ of the \ZZg-graded \CC-algebra $\Huoc{X}$. $\Svec{X}$ forms 
in a natural way a \CC-Lie superalgebra as well as a \ZZg-graded $\Huoc{X}$-module.
In addition one should note, that there is a surjective Lie algebra homomorphism
$\tilde{\beta}_{X}$ from $\Svecg{X}{0}$ to the \CC-Lie algebra $\Vec{\rom{X}}$ of complex 
global vector fields on the body manifold X given by 
\BG
\label{s16c}
\tilde{\beta}_{X}(D)\beta_{X}(f) := \beta_{X}(Df)
\EG
for all $D\in\Svecg{X}{0}$ and all $f\in\Huoc{X}$.\newline
Now let us in particular denote the partial derivatives of~\Hu-functions on the vector superspace $\Vecs{3}{2}$ by $\partial/\partial x^{k}$ in the case of even coordinates and by $\partial/\partial\theta^{\alpha}$ in the case of odd coordinates. If
$N\in\NN^{3}_{0}$ is a multiindex and $K\subseteq\Vecs{3}{2}$ is compact then
\BG
\label{s17}
\semn{f}{3\vert 2}{K,n} := 
\max_{(x^{k},\theta^{\alpha})\in K\atop
\vert N\vert\leq n,M\in\Indns{2}}
\norm{\lp\lp\frac{\partial}{\partial x}\rp^{N}
\lp\frac{\partial}{\partial\theta}\rp^{M}f\rp
(x^{k},\theta^{\alpha})}
\EG
for all $f\in\Huoc{\Vecs{3}{2}}$, where $\vert N\vert$ denotes the length of the multiindex
$N$ and we used the standard notation for partial derivatives of higher order, defines a seminorm on $\Huoc{\Vecs{3}{2}}$. The family of all seminorms for all compact subsets
$K\subseteq\Vecs{3}{2}$ and all natural numbers $n\in\NNN$ induces a locally convex
topology on $\Huoc{\Vecs{3}{2}}$ such that $\Huoc{\Vecs{3}{2}}$ becomes a \ZZg-graded 
\CC-Fr\'{e}chet algebra \cite{Bartocci1}. For our later considerations it is important to note, that the subset $\Poloc{\Vecs{3}{2}}\subseteq\Huoc{\Vecs{3}{2}}$ of all polynomials in the
coordinate projections with complex coefficients forms a dense graded subalgebra of 
$\Huoc{\Vecs{3}{2}}$ \cite{Bartocci1,Kostant1,Treves1}.\newline
The graded ideal $\Idss$ is closed in $\Huoc{\Vecs{3}{2}}$ and consequently $\Huoc{\Ss}$,
endowed with the quotient topology, is also a \ZZg-graded \CC-Fr\'{e}chet algebra. The 
topology is again induced by a family of seminorms, which are given explicitly by
\BG
\label{s18}
\semn{\Quot{f}}{S}{K,n} := \inf_{f\in\Quota{f}}\semn{f}{3\vert 2}{K,n} \, ,
\qquad \Quot{f}\in\Huoc{\Ss} \, .
\EG 
The set of all equivalence classes $\Quot{f}\in\Huoc{\Ss}$
having a polynomial representant forms a dense graded subalgebra of $\Huoc{\Ss}$, which we
denote by $\Poloc{\Ss}$.\newline
$\Cuoc{\RRn{3}}$ and $\Cuoc{\Sp}$ can be endowed with a \CC-Fr\'{e}chet algebra structure
in an analogous way. Then the ``algebraic body maps'' $\beta_{3\vert 2}$ and $\beta_{\Ss}$
are continuous and the subalgebra $\Poloc{\RRn{3}}$ of polynomials in the coordinate projections with complex coefficients as well as its image $\Poloc{\Ss}$ in 
$\Cuoc{\Sp}$ form dense subalgebras in $\Cuoc{\RRn{3}}$ and $\Cuoc{\Sp}$, 
respectively.

Every $f\in\Huoc{\Vecs{3}{2}}$ has by definition a unique ``superfield expansion''
\BG
\label{s19}
f = f_{\emptyset} + f_{4}\theta^{4} + f_{5}\theta^{5} + f_{45}\theta^{4}\theta^{5} \equiv
\sum_{M\in\Indna{2}}f_{M}\theta^{M} \, .
\EG
Inspired by the correspondence of delta functions and Fourier transformation and the rules of Berezin integration \cite{Constantinescu1,DeWitt1}
we define   
\BGG
\label{s20}
\int dxd\theta f(x^{k},\theta^{\alpha})\delta\lp\sum_{i=1}^{3}(x^{i})^{2}+ 2\theta^{4}\theta^{5}-\rho^{2}\rp :=
\qquad \qquad \qquad \qquad \qquad \qquad
\\
= \frac{1}{2}\int\limits_{0}^{\pi}d\vartheta\sin\vartheta\int\limits_{0}^{2\pi}d\varphi\lbr
\frac{\partial\rom{f}_{\emptyset}}{\partial\rho}(\rho,\vartheta,\varphi)+
\frac{1}{\rho}\rom{f}_{\emptyset}(\rho,\vartheta,\varphi)-
\rho\rom{f}_{45}(\rho,\vartheta,\varphi)\rbr \, ,
\nonumber
\EGG
for all $f\in\Huoc{\Vecs{3}{2}}$,
where $\rom{f}_{M}\in\Cuoc{\RRn{3}}$ are the images of $f_{M}\in\Huoc{\Vecs{3}{2}}$ with 
respect to the map (\ref{s13}) expressed in spherical coordinates. (\ref{s20}) induces
a continuous, even \CC-linear map $\Huoc{\Vecs{3}{2}}\longrightarrow\CC$, which vanishes
on $\Idss$ according to the second part of lemma \ref{lemma1s}. Consequently the map
$I:\Huoc{\Ss}\longrightarrow\CC$,
\BG
\label{s21}
I(\Quot{f}) := \int dxd\theta f(x^{k},\theta^{\alpha})\delta\lp\sum_{i=1}^{3}(x^{i})^{2}+ 2\theta^{4}\theta^{5}-\rho^{2}\rp
\, , \qquad f\in\Quot{f} \, ,
\EG
is well-defined and again continuous, even and 
\CC-linear.
In order to introduce an in\-de\-fi\-ni\-te scalar product we define a second ``involution'' 
(besides $*$) $\times:\Huoc{\Vecs{3}{2}}\longrightarrow\Huoc{\Vecs{3}{2}}$ 
via
\BG
\label{s22}
f^{\times} \equiv
\lp\sum_{M\in\Indna{2}}f_{M}\theta^{M}\rp^{\times} :=
f_{\emptyset}^{*} + f_{4}^{*}\theta^{5} - 
f_{5}^{*}\theta^{4} + f_{45}^{*}\theta^{4}\theta^{5} 
\, .
\EG
Apparently (\ref{s22}) is antilinear and fulfills
\BGG
\label{s23}
\lp fg\rp^{\times}&=&(-1)^{\Over{f}\Over{g}}g^{\times}f^{\times}
\nonumber
\\
\lp f^{\times}\rp^{\times}&=&(-1)^{\Over{f}}f 
\EGG
for homogeneous $f,g\in\Huoc{\Vecs{3}{2}}$. Because $\times$ leaves $\Idss$ invariant, 
$\times$ induces the same kind of ``involution'' on $\Huoc{\Ss}$, which we again denote by $\times$. Using (\ref{s21}) as well as $\times$ we can define an even sesquilinear form
on $\Huoc{\Ss}$ by
\BG
\label{s24}
\scalar{\Quotgr{f}}{\Quotgr{g}} := \frac{\rho}{2\pi}I(\Quotgr{f}^{\times}\Quotgr{g}) \, ,
\EG
where the normalization has been chosen such that $\scalar{\Quotgr{1}}{\Quotgr{1}}=1$.
It is continuous in both entries, non-degenerate and fulfills
\BG
\label{s25}
\scalar{\Quotgr{f}}{\Quotgr{g}} = 
\scalar{\Quotgr{g}}{\Quotgr{f}}^{*} \, .
\EG

\section{Action of the (1$\vert$2)-dimensional orthosymplectic Lie superalgebra}

Beside the algebra of functions on a manifold itself there is another important ingredient
for the definition of fuzzy manifolds \cite{Grosse7,Hawkins1,Madore2}:
The action of a Lie group respectively a Lie algebra on the algebra of 
functions.
There are natural graded representations of the (1$\vert$2)-dimensional orthosymplectic Lie superalgebra $\Ort{1}{2}$ on the \ZZg-algebras of \Hu-functions on the 
$(3\vert 2)$-dimensional vector superspace as well as on the (2$\vert$2)-dimensional supersphere. They can be seen as the super-generalizations of the actions of the ordinary angular momentum on the algebras of \Cu-functions on the respective bodies \cite{Daumens2}. These graded representations are reducible. We give their reduction into irreducible subspaces and introduce super-analogues of the spherical harmonics. It should be remarked, that the above mentioned reduction can be found mutatis mutandis also in \cite{Grosse1} and that superspherical harmonics were studied in a different framework in \cite{Daumens2}.

Before analyzing the above mentioned infinite-dimensional graded representations of the
orthosymplectic Lie superalgebra let us first review some well-known facts of this
Lie superalgebra and its finite-dimensional, irreducible graded representations in
order to fix notations and conventions.\newline
The (complex) orthosymplectic Lie superalgebra $\Ort{1}{2}$ is the $(3\vert 2)$-dimensional
\ZZg-graded \CC-vector space spanned by three even basis elements $J_{k}, k=1,2,3,$ and 
two odd basis elements $J_{\alpha}, \alpha=4,5,$ together with the graded Lie bracket
defined by
\BGG
\label{o1}
\Comug{J_{i}}{J_{j}}&:=&i\sum_{k=1}^{3}\varepsilon_{ijk}J_{k}
\nonumber
\\
\Comug{J_{i}}{J_{\alpha}}&:=&\frac{1}{2}
\sum_{\beta=4}^{5}(\sigma_{i})_{\beta\alpha}J_{\beta}
\\
\Comug{J_{\alpha}}{J_{\beta}}&:=&\frac{1}{2}
\sum_{i=1}^{3}(i\sigma_{2}\sigma_{i})_{\alpha\beta}J_{i} \, ,
\nonumber
\EGG
where $\sigma_{k}$ are the Pauli matrices and $\varepsilon_{ijk}$ is 
3-dimensional permutation symbol. It is a basic classical simple Lie superalgebra of type I
\cite{Cornwell1}, whose even part is isomorphic to the 2-dimensional complex special linear Lie algebra $\Sl{2}$ according to (\ref{o1}). The triangular decomposition
\BG
\label{o2}
\Ort{1}{2} = \Oger{N}^{-} \oplus \Oger{H} \oplus \Oger{N}^{+}
\EG
of $\Ort{1}{2}$, where $\Oger{H}$ is the Cartan subalgebra and $\Oger{N}^{\pm}$
are the nilpotent graded Lie subsuperalgebras of $\Ort{1}{2}$ corresponding with the
positive respectively negative roots, is chosen - as usual \cite{Berezin2,Nahm4} - 
according to
\BGG
\label{o3}
\Oger{H}&:=&\rom{span}_{\CCa}\lbr J_{3}\rbr
\nonumber
\\
\Oger{N}^{-}&:=&\rom{span}_{\CCa}\lbr J_{-},J_{5}\rbr
\\
\Oger{N}^{+}&:=&\rom{span}_{\CCa}\lbr J_{+},J_{4}\rbr
\nonumber
\EGG
with
\BG
\label{o4}
J_{\pm} := J_{1} \pm iJ_{2} \, .
\EG
Furthermore there exist two essentially different grade adjoint operations 
$\ddagger_{\lambda}, \lambda=0,1,$ on $\Ort{1}{2}$ \cite{Berezin2,Nahm1,Nahm4}, which correspond, when restricted to the even part of $\Ort{1}{2}$, with the compact real form of $\Ortg{1}{2}{0}$. Explicitly they are given by
\BGG
\label{o5}
J_{i}^{\ddagger_{\lambda}}&:=&J_{i} 
\nonumber
\\
J_{4}^{\ddagger_{\lambda}}&:=&(-1)^{\lambda}J_{5} 
\\
J_{5}^{\ddagger_{\lambda}}&:=&(-1)^{\lambda+1}J_{4} \, . 
\nonumber
\EGG
All irreducible graded representations of $\Ort{1}{2}$ on finite-dimensional \ZZg-graded
\CC-vector spaces are highest weight modules specified by a highest ``weight''
$j\in\frac{1}{2}\NNN$, the so-called superspin, and a definite degree 
$\Over{\varsigma}\in\ZZg$ of the unique, 1-dimensional highest weight space.
Between two irreducible, finite-dimensional graded representations of $\Ort{1}{2}$ 
with the same highest weight $j$ and the same degree $\Over{\varsigma}$ of the highest weight
space there exists an isomorphism of graded $\Ort{1}{2}$-representations, determined by
the requirement, that the highest weight vectors are mapped onto each other 
\cite{Kac1,Kac2}.\newline 
For fixed $j\in\frac{1}{2}\NNN$ and $\Over{\varsigma}\in\ZZg$ let us
denote the corresponding irreducible graded representation by $\cdot^{(j)}$, the
finite-dimensional representation space by $V(j,\Over{\varsigma})$ and by
$e^{(j)}_{j,j,\Over{\varsigma}}\in V(j,\Over{\varsigma})_{\Over{\varsigma}}$
the highest weight vector, specified (up to a complex constant) via
\BGG
\label{o8}
J^{(j)}_{3}e^{(j)}_{j,j,\Over{\varsigma}}&=&je^{(j)}_{j,j,\Over{\varsigma}}
\nonumber
\\
J^{(j)}_{+}e^{(j)}_{j,j,\Over{\varsigma}}&=&
J^{(j)}_{4}e^{(j)}_{j,j,\Over{\varsigma}}= 0
\EGG
according to (\ref{o2}),(\ref{o3}). Then the elements
\BG
\label{o9}
e^{(j)}_{l,m,\Over{\varsigma}+\Over{\mu}} :=
\sqrt{\frac{4^{2(j-l)}\Gamma(l+m+1)}{\Gamma(2j+1)\Gamma(l-m+1)}}\,
J^{(j)\, l-m}_{-}J^{(j)\,\mu}_{5}e^{(j)}_{j,j,\Over{\varsigma}}
\EG
of $V(j,\Over{\varsigma})_{\Over{\varsigma}+\Over{\mu}}$, where the (non-negative)
number $l$ is given by $l:=j-\frac{1}{2}\mu$ and the numbers
$\mu$ and $m$ run through $\lbr 0,1\rbr$ respectively
$\lbr -l,-l+1,\cdots,l-1,l\rbr$, form a homogeneous
basis of $V(j,\Over{\varsigma})$. Using the graded commutation relations
(\ref{o1}) one can deduce the action
\BGG
\label{o10}
J^{(j)}_{3}e^{(j)}_{l,m,\Over{\varsigma}+\Over{\mu}}&=&
m\, e^{(j)}_{l,m,\Over{\varsigma}+\Over{\mu}}
\nonumber
\\
J^{(j)}_{+}e^{(j)}_{l,m,\Over{\varsigma}+\Over{\mu}}&=&
\sqrt{(l-m)(l+m+1)}\, e^{(j)}_{l,m+1,\Over{\varsigma}+\Over{\mu}}
\nonumber
\\
J^{(j)}_{-}e^{(j)}_{l,m,\Over{\varsigma}+\Over{\mu}}&=&
\sqrt{(l+m)(l-m+1)}\, e^{(j)}_{l,m-1,\Over{\varsigma}+\Over{\mu}}
\nonumber
\\
J^{(j)}_{4}e^{(j)}_{j,m,\Over{\varsigma}}&=&
-\frac{1}{2}\sqrt{j-m}\, e^{(j)}_{j-\frac{1}{2},m+\frac{1}{2},\Over{\varsigma}+\Over{1}}
\\
J^{(j)}_{4}e^{(j)}_{j-\frac{1}{2},m,\Over{\varsigma}+\Over{1}}&=&
-\frac{1}{2}\sqrt{j+m+\frac{1}{2}}\, e^{(j)}_{j,m+\frac{1}{2},\Over{\varsigma}}
\nonumber
\\
J^{(j)}_{5}e^{(j)}_{j,m,\Over{\varsigma}}&=&
\frac{1}{2}\sqrt{j+m}\, e^{(j)}_{j-\frac{1}{2},m-\frac{1}{2},\Over{\varsigma}+\Over{1}}
\nonumber
\\
J^{(j)}_{5}e^{(j)}_{j-\frac{1}{2},m,\Over{\varsigma}+\Over{1}}&=&
-\frac{1}{2}\sqrt{j-m+\frac{1}{2}}\, e^{(j)}_{j,m-\frac{1}{2},\Over{\varsigma}}
\nonumber
\EGG
of the homogeneous basis $\lbr J_{3},J_{\pm},J_{\alpha}\rbr$ of $\Ort{1}{2}$
on the homogeneous basis elements (\ref{o9}). (\ref{o9}) and (\ref{o10}) show in 
particular, that the even and the odd subspace of $V(j,\Over{\varsigma})$ correspond to
irreducible $\Sl{2}$-modules with highest weights $l=j$ and $l=j-\frac{1}{2}$ and
\BGG
\label{o11}
&&\mbox{dim}_{\CCa}V(j,\Over{\varsigma})_{\Over{\varsigma}} = 2j + 1 \qquad
\nonumber
\\
&&\mbox{dim}_{\CCa}V(j,\Over{\varsigma})_{\Over{\varsigma}+\Over{1}} = 2j \qquad
\\
&&\mbox{dim}_{\CCa}V(j,\Over{\varsigma}) = 4j + 1 \, . \qquad
\nonumber
\EGG

Now let us introduce the graded derivations
\BGG
\label{o6}
J_{i}^{(3\vert 2)}&:=& 
- i\sum_{j,k=1}^{3}\varepsilon_{ijk}x^{j}\frac{\partial}{\partial x^{k}} +
\frac{1}{2}\sum_{\alpha,\beta=4}^{5}\lp\sigma_{i}\rp_{\alpha\beta}
\theta^{\alpha}\frac{\partial}{\partial\theta^{\beta}} \, ,
\qquad i=1,2,3 \, ,
\nonumber
\\
J_{\alpha}^{(3\vert 2)}&:=& 
\frac{1}{2}\sum_{k=1}^{3}\sum_{\beta=4}^{5}\lbr
ix^{k}\lp\sigma_{2}\sigma_{k}\rp_{\alpha\beta}
\frac{\partial}{\partial\theta^{\beta}}-
\theta^{\beta}\lp\sigma_{k}\rp_{\beta\alpha}
\frac{\partial}{\partial x^{k}}\rbr \, ,
\qquad \alpha=4,5 \, ,
\EGG
of $\Huoc{\Vecs{3}{2}}$. The corresponding \CC-linear map 
$\cdot^{(3\vert 2)}:\Ort{1}{2}\longrightarrow\Svec{\Vecs{3}{2}}$ 
is a graded re\-pre\-sen\-ta\-tion of $\Ort{1}{2}$, that leaves $\Idss$ (according to
lemma \ref{lemma1s}) invariant. Consequently, defining the action via the action on representants, $\cdot^{(3\vert 2)}$ induces a graded representation
$\cdot^{(S)}:\Ort{1}{2}\longrightarrow\Svec{\Ss}$ and the latter is a ``grade star representation with respect to $\ddagger_{1}$'',
that is, 
\BG
\label{o7}
\scalar{\Quot{f}}{J_{A}^{(S)}\Quot{g}} =
(-1)^{\Over{J}_{\! A}\Over{\Quota{f}}}\scalarl{J_{A}^{\ddagger_{1}(S)}\Quot{f}\,}{\Quot{g}}
\EG
is fulfilled for all homogeneous $\Quot{f},\Quot{g}\in\Huoc{\Ss}$ and 
$A=1,\cdots,5$.\newline
$\cdot^{(3\vert 2)}$ as well as $\cdot^{(S)}$ are reducible. In order to find their
decomposition into irreducible subspaces we note, that the even polynomials
\BG
\label{o12}
Y^{s,j}_{j,j,\Over{0}} := \frac{\sqrt{\Gamma(2j+1)}}{2^{j}\Gamma(j+1)\rho^{j+2s}}
\lp\sum_{i=1}^{3}(x^{k})^{2}+2\theta^{4}\theta^{5}\rp^{s}\lp x^{1}+ix^{2}\rp^{j} \, ,
\EG
for all $j,s\in\NNN$, and the odd polynomials
\BG
\label{o13}
Y^{s,j}_{j,j,\Over{1}} := \frac{\sqrt{\Gamma(2j+1)}}
{2^{j-\frac{1}{2}}\Gamma(j+\frac{1}{2})\rho^{j+\frac{3}{2}+2s}}
\lp\sum_{i=1}^{3}(x^{k})^{2}+2\theta^{4}\theta^{5}\rp^{s}
\lp x^{1}+ix^{2}\rp^{j-\frac{1}{2}}\lp
x^{3}\theta^{4}+(x^{1}+ix^{2})\theta^{5}\rp \, ,
\EG
for all $j\in\NNN+\frac{1}{2},s\in\NNN$, are highest weight vectors of $\cdot^{(3\vert 2)}$
with highest weight $j$. Such highest weight vectors with different $s$ but the same $j$ are
mapped under the canonical projection onto the same element of $\Poloc{\Ss}$, which we will
denote by $\Sphar{j}{j}{j}{2j}$. The normalization in (\ref{o12}) and (\ref{o13}) is
chosen such that
\BG
\label{o14}
\scalar{\Sphar{j}{j}{j}{2j}}{\Sphar{j}{j}{j}{2j}} = 1
\EG
is fulfilled. As we will see, the graded $\Ort{1}{2}$-submodules
\BGG
\label{o15}
V^{s,j}&:=&U(\Ort{1}{2})^{(3\vert 2)}Y^{s,j}_{j,j,\Over{2j}}
\nonumber
\\
\VSphar{j}&:=&U(\Ort{1}{2})^{(S)}\Sphar{j}{j}{j}{2j} \, ,
\EGG
where $U(\Ort{1}{2})$ denotes the enveloping algebra of $\Ort{1}{2}$, constitute the 
irreducible direct summands of the graded representations $\cdot^{(3\vert 2)}$ and $\cdot^{(S)}$.
\begin{lemma} 
\label{lemma1o}
The restriction of $\cdot^{(3\vert 2)}$ to $V^{s,j}$ is an irreducible graded
representation with highest weight vector $Y^{s,j}_{j,j,\Over{2j}}$, highest weight $j$
and degree $\Over{2j}$ of the highest weight vector. 
Moreover the dense graded sub\-al\-ge\-bra $\Poloc{\Vecs{3}{2}}\subseteq\Huoc{\Vecs{3}{2}}$ can be decomposed as \ZZg-graded $\Ort{1}{2}$-module according to
\BG 
\label{o16}
\Poloc{\Vecs{3}{2}} = \bigoplus_{s\in\NNNs\atop j\in\frac{1}{2}\NNNs}V^{s,j} \, .
\EG
\end{lemma}
${\sl Proof}$: Let us assume, that the restriction of $\cdot^{(3\vert 2)}$ to $V^{s,j}$ is
reducible. Then there exists a graded subrepresentation of $\cdot^{(3\vert 2)}$ on a graded
vector subspace $W\subset V^{s,j}$ and a non-trivial homogeneous 
$w\in W\setminus V^{s,j}$. Because $\cdot^{(3\vert 2)}$ is standard-cyclic, $w$ is given 
either by
\BD
w = \sum_{q=0}^{N_{j}}\alpha_{q}J_{-}^{(3\vert 2)\, q}Y^{s,j}_{j,j,\Over{2j}} \, ,
\qquad \alpha_{q}\in\CC, \alpha_{N_{j}}\neq 0 \, ,
\ED
or by
\BD
w = \sum_{q=0}^{N_{j}}
\alpha_{q}J_{-}^{(3\vert 2)\, q}J_{5}^{(3\vert2)}Y^{s,j}_{j,j,\Over{2j}} \, ,
\qquad \alpha_{q}\in\CC, \alpha_{N_{j}}\neq 0 \, .
\ED
Using the explicit expressions (\ref{o6}),(\ref{o12}) and (\ref{o13}) one finds by induction
$N_{j}\leq 2j$ and
\BD
J_{+}^{(3\vert 2)\, N_{j}}w = \gamma_{j}Y^{s,j}_{j,j,\Over{2j}} \, ,
\qquad \gamma_{j}\in\CC\setminus\{ 0\} \, ,
\ED
in the first case and $N_{j}\leq 2j-1$ and
\BD
J_{+}^{(3\vert 2)\, N_{j}}J_{4}^{(3\vert 2)}w = \gamma_{j}Y^{s,j}_{j,j,\Over{2j}} \, ,
\qquad \gamma_{j}\in\CC\setminus\{ 0\} \, ,
\ED
in the second case. In any case one has $Y^{s,j}_{j,j,\Over{2j}}\in W$ implying
the contradiction $W=V^{s,j}$.\newline
Apparently $\cdot^{(3\vert 2)}$ does not only leave $\Poloc{\Vecs{3}{2}}$ invariant but also
the graded vector subspaces $\Polocn{n}{\Vecs{3}{2}}\subseteq\Poloc{\Vecs{3}{2}}$ of polynomials in the coordinate projections with complex coefficients of degree $n\in\NNN$.
Consequently we have $V^{s,j}\subseteq\Polocn{j+\frac{3}{2}\Over{2j}+2s}{\Vecs{3}{2}}$,
where the summand $\frac{3}{2}\Over{2j}$ is of course to be understood as 0, if $j\in\NNN$,
and as $\frac{3}{2}$, if $j\in\NNN+\frac{1}{2}$. In order to show (\ref{o16}) we only have
to prove
\BG
\label{o17}
\Polocn{n}{\Vecs{3}{2}} = \bigoplus_{j+\frac{3}{2}\Over{2j}+2s=n}V^{s,j}
\EG
for all $n\in\NNN$. Using the fact, that highest weight vectors of $V^{s,j}$ and $V^{s',j'}$,
with $s\neq s',j\neq j'$, such that $j+\frac{3}{2}\Over{2j}+2s=j'+\frac{3}{2}\Over{2j}'+2s'=n$,
have to be linear independent because they are highest weight vectors to different highest
weights, one finds, that the intersection of $V^{s,j}$ with the sum of all other vector
spaces $V^{s',j'}$ with $s\neq s',j\neq j'$ such that $j+\frac{3}{2}\Over{2j}+2s=j'+\frac{3}{2}\Over{2j}'+2s'=n$ is $\{ 0\}$.
Consequently $\oplus_{j+\frac{3}{2}\Over{2j}+2s=n}V^{s,j}$ is a well-defined graded vector
subspace of $\Polocn{n}{\Vecs{3}{2}}$. In order to show equality it is enough to show the
equality of dimensions. But using (\ref{o11}) we find $2n^{2}+2n+1$ for
$\mbox{dim}_{\CCa}\lp\oplus_{j+\frac{3}{2}\Over{2j}+2s=n}V^{s,j}\rp$ by induction, which is
exactly $\mbox{dim}_{\CCa}\Polocn{n}{\Vecs{3}{2}}$.\hfill $\Box$ 

\vspace{2mm}

Although the above result about the representation $\cdot^{(3\vert 2)}$ is not of primary interest for itself it is of central importance for deducing the following reduction of
$\cdot^{(S)}$.
\begin{proposition}
\label{theo1o}
The restriction of $\cdot^{(S)}$ to $\VSphar{j}$ is an irreducible graded
representation with highest weight vector $\Sphar{j}{j}{j}{2j}$, highest weight $j$
and degree $\Over{2j}$ of the highest weight vector. 
Moreover the dense graded sub\-al\-ge\-bra $\Poloc{\Ss}\subseteq\Huoc{\Ss}$ can be decomposed as \ZZg-graded $\Ort{1}{2}$-module according to
\BG 
\label{o18}
\Poloc{\Ss} = \bigoplus_{j\in\frac{1}{2}\NNNa}\VSphar{j} \, .
\EG
\end{proposition}
${\sl Proof}:$ The first statement is a consequence of the facts, that $\Sphar{j}{j}{j}{2j}$
is non-trivial and that the canonical projection $\Huoc{\Vecs{3}{2}}\longrightarrow
\Huoc{\Ss}$ is an surjective, even homomorphism of $\Ort{1}{2}$-modules.\newline 
Because of lemma \ref{lemma1o} we only have to show, that the intersection of $\VSphar{j}$
and the sum of all other vector subspaces $\VSphar{j'}$ is trivial, in order to prove (\ref{o18}).  But this is an easy consequence of the fact, that $\VSphar{j}$
and $\VSphar{j'}$ are irreducible \ZZg-graded $\Ort{1}{2}$-modules with different highest 
weights. \hfill $\Box$ 

\vspace{2mm}

According to proposition \ref{theo1o} and (\ref{o9}) the elements of 
$\VSphar{j}\subseteq\Poloc{\Ss}\subseteq\Huoc{\Ss}$
\BG
\label{o19}
\Sphard{j}{l}{m}{2j}{\mu} :=
\sqrt{\frac{4^{2(j-l)}\Gamma(l+m+1)}{\Gamma(2j+1)\Gamma(l-m+1)}}\,
J^{(S)\, l-m}_{-}J^{(S)\,\mu}_{5}\Sphar{j}{j}{j}{2j} \, ,
\EG
with $l:=j-\frac{1}{2}\mu$ and $\mu\in\lbr 0,1\rbr$, $m\in\lbr -l,-l+1,\cdots,l-1,l\rbr$,  form a homogeneous basis of $\VSphar{j}$ for all $j\in\frac{1}{2}\NNN$. We will call them
superspherical harmonics. Of course, the action of $\Ort{1}{2}$ on the superspherical
harmonics is given by (\ref{o10}) and using (\ref{o7}) and (\ref{o14}) we find, that they
are (pseudo)orthonormalized according to
\BG
\label{o20}
\scalar{\Sphard{j}{l}{m}{2j}{\mu}}{\Sphard{j'}{l'}{m'}{2j'}{\mu'}} =
(-1)^{\delta_{\Over{2j}\,\Over{1}}\delta_{\mu 1}}
\delta_{jj'}\delta_{\mu\mu'}\delta_{mm'} \, .
\EG
  
\section{Truncation and the fuzzy supersphere}

We truncate the direct sum (\ref{o18}) of graded subrepresentations of $\Ort{1}{2}$ on the
\ZZg-graded algebra of complex-Grassmann-valued \Hu-functions on the $(2\vert 2)$-dimensional
supersphere at each integer superspin and establish a new, \ZZg-graded associative product on each of these truncated sums. In this way we get the fuzzy supersphere, that is a whole sequence of finite-dimensional, noncommutative \ZZg-graded algebras possessing a ``graded-commutative limit''. The procedure described above was studied in a similar way in \cite{Grosse1}. Here we have slightly different conventions and we will discuss the 
graded-commutative limit in greater detail, which includes the introduction of noncommutative superspherical harmonics in particular.
 
Let us first formulate our aim, that is the basic idea of the construction of fuzzy 
manifolds applied to our case. For every $q\in\frac{1}{2}\NN$ let us introduce the truncated direct sum
\BG
\label{t1}
{\cal H}_{q} := \bigoplus_{j\in\frac{1}{2}\NNNs\atop j\leq q}\VSphar{j}
\EG 
of \ZZg-graded $\Ort{1}{2}$-modules. For some infinite subset $\SUB\subseteq\frac{1}{2}\NNN$
and every $q\in\SUB$ we want to find \ZZg-graded \CC-algebras ${\cal A}_{q}$, which are at
the same time \ZZg-graded $\Ort{1}{2}$-modules together with even $\Ort{1}{2}$-module
isomorphisms $\psi_{q}:{\cal H}_{q}\longrightarrow{\cal A}_{q}$. Denoting by $\iota_{q'q}, q,q'\in\SUB, q\leq q',$ the canonical injections ${\cal H}_{q}\longrightarrow{\cal H}_{q'}$ we can introduce even, injective $\Ort{1}{2}$-module homomorphisms 
$\eta_{q'q}:{\cal A}_{q}\longrightarrow{\cal A}_{q'}$ by
\BG
\label{t15}
\eta_{q'q} := \psi_{q'}\circ\iota_{q'q}\circ\psi_{q}^{-1} \, ,
\EG
which fulfill
\BG
\label{t16}
\eta_{q''q'}\circ\eta_{q'q} = \eta_{q''q}
\EG
for all $q,q',q''\in\SUB, q\leq q'\leq q''$. Consequently 
$({\cal H}_{q},\iota_{q'q})$ and $({\cal A}_{q},\eta_{q'q})$ are isomorphic directed 
systems of \ZZg-graded $\Ort{1}{2}$-modules and their direct limits can be identified
with $\Poloc{\Ss}$ (as \ZZg-graded $\Ort{1}{2}$-modules). The corresponding homomorphisms
${\cal H}_{q}\longrightarrow\Poloc{\Ss}$ and ${\cal A}_{q}\longrightarrow\Poloc{\Ss}$
of \ZZg-graded $\Ort{1}{2}$-modules are denoted by $\iota_{q}$ and $\eta_{q}$,
respectively.\newline
Now let $\Quotgr{f}$ and $\Quotgr{f}\,'$ be elements of $\Poloc{\Ss}$. Then there will be a
number $p\in\SUB$ such that (omitting the canonical injections $\iota_{p}$ and
$\iota_{qp}$ as we will always do in the sequel) $\Quotgr{f},\Quotgr{f}\,'\in{\cal H}_{q}$
for all $q\geq p$. We cannot form products $\Quotgr{f}\Quotgr{f}\,'$ in the truncated sums 
${\cal H}_{q}$, but we can form $\psi_{q}(\Quotgr{f})\psi_{q}(\Quotgr{f}\,')$ in the isomorphic objects ${\cal A}_{q}$. A priori these products are not connected with the product
$\Quotgr{f}\Quotgr{f}\,'\in\Huoc{\Ss}$ in any way, but we can map them
into the direct limit according to
\BG
\label{t17}
\lp\Quotgr{f}\Quotgr{f}\,'\rp_{q} := \eta_{q}\lp\psi_{q}(\Quotgr{f})\psi_{q}(\Quotgr{f}\,')\rp =
\psi_{q}^{-1}\lp\psi_{q}(\Quotgr{f})\psi_{q}(\Quotgr{f}\,')\rp \in \Poloc{\Ss} \subseteq \Huoc{\Ss}
\EG
and ``compare'' $(\Quotgr{f}\Quotgr{f}\,')_{q}$ with $\Quotgr{f}\Quotgr{f}\,'$. More exactly
$\{(\Quotgr{f}\Quotgr{f}\,')_{q}\}_{p\leq q,q\in\SUBa}$ is a sequence in
$\Huoc{\Ss}$, whose convergence to $\Quotgr{f}\Quotgr{f}\,'$ (with respect to the Fr\'{e}chet topology) can be investigated.
\begin{definition} 
\label{def1t}
The directed system $({\cal A}_{q},\eta_{q'q})$ is said to possess a graded-commutative limit
if
\BG
\label{t18}
\lim_{q\rightarrow\infty}\lp\Quotgr{f}\Quotgr{f}\,'\rp_{q} =
\Quotgr{f}\Quotgr{f}\,'
\EG
is fulfilled for all $\Quotgr{f},\Quotgr{f}\,'\in\Poloc{\Ss}$.
\end{definition}
Because of the algebraic structure of each ${\cal A}_{q}$, we can view each ${\cal A}_{q}$
as ``noncommutative \Hu-supermanifold''; the existence of the graded-commutative limit
guarantees the relation with the $(2\vert 2)$-dimensional supersphere. We will show
in the sequel, that a directed system with graded-commutative limit really exists.

We choose $\SUB=\NN$. For all $q\in\NN$ one finds by induction 
\BGG
\label{t2}
&& \mbox{dim}_{\CCa}{\cal H}_{q,\Over{0}} = q^{2} + \lp q+1\rp^{2} \qquad
\nonumber
\\
&& \mbox{dim}_{\CCa}{\cal H}_{q,\Over{1}} = 2q\lp q+1\rp \qquad
\\
&& \mbox{dim}_{\CCa}{\cal H}_{q} = \lp 2q+1\rp^{2} \, , \qquad
\nonumber
\EGG
from which we can conclude, that there will exist isomorphisms of \ZZg-graded \CC-vector
spaces between ${\cal H}_{q}$ and the \ZZg-graded \CC-algebra $\End{V(\frac{q}{2},\Over{1})}$
of all endomorphisms of the \ZZg-graded representation space of the irreducible graded 
$\Ort{1}{2}$-representation with highest weight $\frac{q}{2}$ and odd highest weight 
vector.\newline
On each of the \ZZg-graded representation spaces $V(\frac{q}{2},\Over{1})$ we can
introduce a scalar product by
\BG
\label{t3}
\scalar{e^{(\frac{q}{2})}_{l,m,\Over{1}+\Over{\mu}}}
{e^{(\frac{q}{2})}_{l',m',\Over{1}+\Over{\mu'}}} =
\delta_{\mu\mu'}\delta_{mm'}
\EG
for all $\mu,\mu'\in\{ 0,1\}$, $m\in\{-l,\cdots,l\}$, $m'\in\{-l',\cdots,l'\}$, such that $V(\frac{q}{2},\Over{1})$ becomes a \ZZg-graded Hilbert space. With respect to this scalar product and the grade adjoint operation $\ddagger_{0}$, the irreducible graded representation $\cdot^{(\frac{q}{2})}$ is a grade star representation \cite{Berezin2,Nahm4}.
Employing the superadjoint operation $\ddagger$ with respect to this scalar product as well as the supertrace $\rom{Tr}_{s}$ we can define a sesquilinear form $\scalarK{\cdot}{\cdot}{\frac{q}{2}}:\End{V(\frac{q}{2},\Over{1})}\times\End{V(\frac{q}
{2},\Over{1})}\longrightarrow\CC$ via
\BG
\label{t5}
\scalarK{f}{g}{\frac{q}{2}} := -\rom{Tr}_{s}\lp f^{\ddagger}g\rp =
\sum_{\mu=0}^{1}\sum_{m=-l}^{l}(-1)^{\mu}
\scalar{e^{(\frac{q}{2})}_{l,m,\Over{1}+\Over{\mu}}}
{f^{\ddagger}g\lp e^{(\frac{q}{2})}_{l,m,\Over{1}+\Over{\mu}}\rp} \, .
\EG
The normalization has been chosen such that $\langle\rom{Id}_{V(q/2,\Over{1})}\vert\rom{Id}_{V(q/2,\Over{1})}\rangle_{\frac{q}{2}}=1$.
$\scalarK{\cdot}{\cdot}{\frac{q}{2}}$ is even, non-de\-ge\-ne\-rate and fulfills
\BG
\label{t6}
\scalarK{f}{g}{\frac{q}{2}} = \scalarK{g}{f}{\frac{q}{2}}^{*} \, .
\EG
Beside this indefinite scalar product, which has exactly the same properties as (\ref{s24}), we can also establish a \ZZg-graded $\Ort{1}{2}$-module structure on $\End{V(\frac{q}{2},\Over{1})}$ in a natural way by defining  
$\rom{ad}^{(\frac{q}{2})}:\Ort{1}{2}\longrightarrow\Derc
{\End{V(\frac{q}{2},\Over{1})}}\subseteq\Plv{\End{V(\frac{q}{2},\Over{1})}}$ via
\BG
\label{t7}
\rom{ad}^{(\frac{q}{2})}J(f) := \Comug{J^{(\frac{q}{2})}}{f}
\EG
for all $J\in\Ort{1}{2}$, $f\in\End{V(\frac{q}{2},\Over{1})}$. Here we denoted by $\Plv{\End{V(\frac{q}{2},\Over{1})}}$ the general linear Lie superalgebra of the \ZZg-graded
\CC-vector space $\End{V(\frac{q}{2},\Over{1})}$ and by $\Comug{\cdot}{\cdot}$ its graded Lie
bracket. Using the fact, that $\cdot^{(\frac{q}{2})}$ is a grade star representation with respect to $\ddagger_{0}$, we can deduce, that $\rom{ad}^{(\frac{q}{2})}$ is a grade
star representation with respect to $\ddagger_{1}$.\newline
The graded representations $\rom{ad}^{(\frac{q}{2})}$ are reducible for all $q\in\NN$. In
order to find their reduction into irreducible subspaces we proceed as we did for
the infinite-dimensional case. Employing (\ref{o10}) and the graded Leibniz rule we can check, that the even endomorphisms
\BG
\label{t8}
\NSphar{\frac{q}{2}}{j}{j}{j}{0} := 
\sqrt{\frac{2^{j}(2j-1)!!\Gamma(q-j+1)}{\Gamma(j+1)\Gamma(q+j+1)}}
J_{+}^{(\frac{q}{2})\, j} \, ,
\EG
for all $j\in\NNN, j\leq q$, as well as the odd endomorphisms
\BG
\label{t9}
\NSphar{\frac{q}{2}}{j}{j}{j}{1} := \frac{1}{q+\frac{1}{2}}
\sqrt{\frac{2^{j+\frac{7}{2}}(2j)!!\Gamma(q-j+\frac{1}{2})}
{\Gamma(j+\frac{1}{2})\Gamma(q+j+\frac{3}{2})}}J_{+}^{(\frac{q}{2})\, j-\frac{1}{2}}\lb\lp
J_{3}^{(\frac{q}{2})}-\frac{3}{4}\rp J_{4}^{(\frac{q}{2})}+
J_{+}^{(\frac{q}{2})}J_{5}^{(\frac{q}{2})}\rb \, ,
\EG
for all $j\in\NNN+\frac{1}{2}, j\leq q$, are highest weight vectors of $\rom{ad}^{(\frac{q}{2})}$ with highest weight $j$ and normalized according to
\BG
\label{t10}
\scalar{\NSphar{\frac{q}{2}}{j}{j}{j}{2j}}{\NSphar{\frac{q}{2}}{j}{j}{j}{2j}}_{\frac{q}{2}} = 
1 \, .
\EG
The corresponding graded $\Ort{1}{2}$-submodules
\BG
\label{t11}
\NVSphar{\frac{q}{2}}{j} := \rom{ad}^{(\frac{q}{2})}\lp U(\Ort{1}{2})\rp\NSphar{\frac{q}{2}}{j}{j}{j}{2j} 
\EG
are the direct summands we are looking for.
\begin{proposition}
\label{theo1t}
The restriction of $\rom{ad}^{(\frac{q}{2})}$ to $\NVSphar{\frac{q}{2}}{j}$ is an irreducible graded representation of $\Ort{1}{2}$ with highest weight vector $\NSphar{\frac{q}{2}}{j}{j}{j}{2j}$, highest weight $j$ and degree $\Over{2j}$ of the highest weight vector. Moreover $\End{V(\frac{q}{2},\Over{1})}$ can be decomposed as \ZZg-graded $\Ort{1}{2}$-module according to
\BG
\label{t12}
\End{V(\frac{q}{2},\Over{1})} = 
\bigoplus_{j\in\frac{1}{2}\NNNs\atop j\leq q}\NVSphar{\frac{q}{2}}{j} \, . 
\EG
\end{proposition}
${\sl Proof}:$ The proof of irreducibility can be taken over literally from
lemma \ref{lemma1o}: The numbers $N_{j}$ can again be determined by induction (with the
same result as in lemma \ref{lemma1o}) but now by using the explicit expressions
(\ref{t8}),(\ref{t9}). The fact, that we are considering only graded highest weight
modules with different highest weights guarantees that the direct sum in (\ref{t12}) is
well-defined and the equality with $\End{V(\frac{q}{2},\Over{1})}$ results from our dimensional considerations (\ref{t2}). \hfill $\Box$ 

\vspace{2mm}

Consequently for every $q\in\NN$ and $j\leq q$ the elements
\BG
\label{t13}
\NSphard{\frac{q}{2}}{j}{l}{m}{2j}{\mu} :=
\sqrt{\frac{4^{2(j-l)}\Gamma(l+m+1)}{\Gamma(2j+1)\Gamma(l-m+1)}}
\lp\rom{ad}^{(\frac{q}{2})}J_{-}\rp^{l-m}\lp
\rom{ad}^{(\frac{q}{2})}J_{5}\rp^{\mu}\NSphar{\frac{q}{2}}{j}{j}{j}{2j} 
\EG
of $\NVSphar{\frac{q}{2}}{j}$, with $l:=j-\frac{1}{2}\mu$ and $\mu\in\lbr 0,1\rbr$, 
$m\in\lbr -l,-l+1,\cdots,l-1,l\rbr$, form a homogeneous basis of $\NVSphar{\frac{q}{2}}{j}$.
Because of the isomorphisms (\ref{t19}) defined below, we will call these elements of $\End{V(\frac{q}{2},\Over{1})}$ noncommutative superspherical harmonics and an analogous calculation which yielded (\ref{o20}) shows, that they are again (pseudo)orthonormalized according to
\BG
\label{t14}
\scalarK{\NSphard{\frac{q}{2}}{j}{l}{m}{2j}{\mu}}
{\NSphard{\frac{q}{2}}{j'}{l'}{m'}{2j'}{\mu'}}{\frac{q}{2}} =
(-1)^{\delta_{\Over{2j}\,\Over{1}}\delta_{\mu 1}}
\delta_{jj'}\delta_{\mu\mu'}\delta_{mm'} \, .
\EG
In consideration of proposition \ref{theo1t} we can introduce lots of 
even $\Ort{1}{2}$-module isomorphisms 
$\psi_{q}:{\cal H}_{q}\longrightarrow\End{V(\frac{q}{2},\Over{1})}$ and we choose
especially for every $q\in\NN$ the \CC-linear map defined by
\BG
\label{t19}
\psi_{q}(\Sphard{j}{l}{m}{2j}{\mu}) := \NSphard{\frac{q}{2}}{j}{l}{m}{2j}{\mu}
\EG
for all $j\leq q, \mu\in\lbr 0,1\rbr$ and $m\in\lbr -l,-l+1,\cdots,l-1,l\rbr$, because
the corresponding directed system $(\End{V(\frac{q}{2},\Over{1})},\eta_{q'q})$ has the
desired property.
\begin{proposition}
\label{theo2t}
The directed system $(\End{V(\frac{q}{2},\Over{1})},\eta_{q'q})$ corresponding with
$(\ref{t19})$ possesses a graded-commutative limit.
\end{proposition}
${\sl Proof}:$ Because the topology on $\Huoc{\Ss}$ is induced by seminorms (\ref{s18}),
superspherical harmonics - ordinary and noncommutative as well - are (pseudo)orthonormalized
and $\cdot^{(S)}$ and $\rom{ad}^{(\frac{q}{2})}$ are grade star representations of the same type, it is enough to show
\BG
\label{t20}
\lim_{q\rightarrow\infty}\scalarK{\NSphard{\frac{q}{2}}{j}{l}{m}{2j}{\mu}}
{\NSphard{\frac{q}{2}}{j'}{l'}{m'}{2j'}{\mu'}\NSphard{\frac{q}{2}}
{j''}{l''}{m''}{2j''}{\mu''}}{\frac{q}{2}} = 
\scalar{\Sphard{j}{l}{m}{2j}{\mu}}{\Sphard{j'}{l'}{m'}{2j'}{\mu'}\Sphard{j''}{l''}
{m''}{2j''}{\mu''}}
\EG
for all $j',j''\in\frac{1}{2}\NNN,\mu',\mu''\in\{ 0,1\},m'\in\{-l',\cdots,l'\},
m'\in\{-l',\cdots,l'\}$ and $j\in\frac{1}{2}\NNN,\mu\in\{ 0,1\},m\in\{-l,\cdots,l\}$
with $\vert j'-j''\vert\leq j\leq j'+j'',m=m'+m'',\Over{\mu}=
\Over{2j+2j'+2j''+\mu'+\mu''}$.\newline
Now we can interpret $\Sphard{j'}{l'}{m'}{2j'}{\mu'}$ and $\NSphard{\frac{q}{2}}{j'}{l'}{m'}{2j'}{\mu'}$ as multiplication operators in 
$\Poloc{\Ss}$ and $\End{V(\frac{q}{2},\Over{1})}$, respectively. Then the sets
$\{~\Sphard{j'}{l'}{m'}{2j'}{\mu'}~\vert~\mu'=0,1; m'=-l',\cdots,l'~\}$ and 
$\{~\NSphard{\frac{q}{2}}{j'}{l'}{m'}{2j'}{\mu'}~\vert~\mu'=0,1; m'=-l',\cdots,l'~\}$ are
irreducible $\Ort{1}{2}$-tensor operators and we can apply the
$\Ort{1}{2}$-Wigner-Eckart theorem \cite{Mezincescu1,Zeng1} to conclude, that we can restrict our attention to the cases $\mu'=\mu''=0,m'=j',m''=j''$.\newline 
We find
\BD
\Sphar{j'}{j'}{j'}{2j'}\Sphar{j''}{j''}{j''}{2j''} = c_{j'j''}\Sphar{j'+j''}{j'+j''}{j'+j''}{2(j'+j'')} \, , 
\qquad j',j''\in\frac{1}{2}\NNN,
\ED
as well as
\BD
\NSphar{\frac{q}{2}}{j'}{j'}{j'}{2j'}\NSphar{\frac{q}{2}}{j''}{j''}{j''}{2j''} = c_{j'j''}^{q}\NSphar{\frac{q}{2}}{j'+j''}{j'+j''}{j'+j''}{2(j'+j'')} \, , 
\qquad j',j''\in\frac{1}{2}\NNN,
\ED
where the coefficients $c_{j'j''},c_{j'j''}^{q}\in\RR$, which are also determined by the explicit expressions (\ref{o12}),(\ref{o13}) and (\ref{t8}),(\ref{t9}), fulfill
\BD
\lim_{q\rightarrow\infty}c_{j'j''}^{q} = c_{j'j''}
\ED 
for all $j',j''\in\frac{1}{2}\NNN$. But because of (\ref{o20}) and (\ref{t14})
this proves the proposition.
\hfill $\Box$ 

\vspace{2mm}

Consequently we have succeeded in finding a sequence of \ZZg-graded \CC-algebras tending
in the limit described above to the \ZZg-graded \CC-algebra $\Huoc{\Ss}$ of \Hu-functions 
on the $(2\vert 2)$-dimensional supersphere $\Ss$. Transposed to the
language of noncommutative geometry this means, that we have approximated the
$(2\vert 2)$-dimensional supersphere by a sequence of noncommutative supermanifolds.
Because our construction is exactly the same as the one of ordinary fuzzy manifolds
\cite{Grosse7,Hawkins1,Madore2} we will call the directed system
$(\End{V(\frac{q}{2},\Over{1})},\eta_{q'q})$ ($(2\vert 2)$-dimensional) fuzzy supersphere.
Each one of the \ZZg-graded \CC-algebras and graded $\Ort{1}{2}$-modules 
$\End{V(\frac{q}{2},\Over{1})}$ we will call truncated ($(2\vert 2)$-dimensional) 
supersphere and we introduce the shorter notation $\NSs{q}$ for it.\newline
It is worthwhile to mention the following nice property of the
isomorphisms $\psi_{q}$ and the truncated superspheres $\NSs{q}$: 
For all $q\in\NN$ we find
\BGG
\label{t21}
&& \psi_{q}(\Quotgr{x}^{\, k}) = \frac{2\rho}{\sqrt{q(q+1)}}J^{(\frac{q}{2})}_{k} =:
X^{(\frac{q}{2})}_{k} \, , \qquad k=1,2,3, \qquad
\nonumber
\\
&& \psi_{q}(\Quotgr{\theta}^{\alpha}) = 
\frac{2\rho}{\sqrt{q(q+1)}}J^{(\frac{q}{2})}_{\alpha} =:
\Theta^{(\frac{q}{2})}_{\alpha} \, , \qquad \alpha=4,5, \qquad
\EGG
and consequently
\BGG
\label{t22}
\sum_{k=1}^{3}\psi_{q}(\Quotgr{x}^{\, k})^{2} + 
\psi_{q}(\Quotgr{\theta}^{4})\psi_{q}(\Quotgr{\theta}^{5}) -
\psi_{q}(\Quotgr{\theta}^{5})\psi_{q}(\Quotgr{\theta}^{4}) \equiv \qquad \qquad
\nonumber
\\
\equiv
\sum_{k=1}^{3}\lp X^{(\frac{q}{2})}_{k}\rp^{2} +
\Theta^{(\frac{q}{2})}_{4}\Theta^{(\frac{q}{2})}_{5} -
\Theta^{(\frac{q}{2})}_{5}\Theta^{(\frac{q}{2})}_{4} =
\rho^{2}\rom{Id}_{V(\frac{q}{2},\Over{1})} \, ,
\EGG
because the left hand side is $4\rho^{2}/(q(q+1))$ times the representation of the
standard second-order Casimir operator \cite{Nahm4} of $\Ort{1}{2}$. That is, the
defining relation of the $(2\vert 2)$-dimensional supersphere is fulfilled on each truncated
supersphere.

\section{The noncommutative body map}

The existence of a body and an ``algebraic body map'' is of great structural importance in the
theory of supermanifolds respectively graded manifolds and one should have an fuzzy analogue. We will see, that it is very natural to interpret the fuzzy sphere as ``noncommutative body'' of the fuzzy supersphere. The corresponding noncommutative body map will be a surjective homomorphism of directed systems, but none of the single maps will be an algebra homomorphism (which would be impossible as map between simple algebras of different dimensions). The algebra homomorphism property will be recovered in the ``graded-commutative limit''.

Let us first describe the ``noncommutative body'' of the fuzzy supersphere, that is the
fuzzy sphere, in an adequate language (see also \cite{Madore2,Madore10,Grosse5,Grosse1}). 
The \CC-linear map 
$\cdot^{(3)}:\Sl{2}\longrightarrow\Vec{\RRn{3}}$, defined by
\BG
\label{b1}
\rom{J}^{(3)}_{i} := -i\sum_{j,k=1}^{3}\varepsilon_{ijk}\rom{x}^{j}\partial_{k} \, ,
\qquad i=1,2,3,
\EG
is a representation of $\Sl{2}$. It leaves $\Idsp$ invariant and induces a representation 
$\cdot^{(\rom{S})}:\Sl{2}\longrightarrow\Vec{\Sp}$, which is a star representation with
respect to the adjoint operation $\dagger$ corresponding with the compact real form
of $\Sl{2}$ and the scalar product 
$\scalar{\cdot}{\cdot}:\Cuoc{\Sp}\times\Cuoc{\Sp}\longrightarrow\CC$, defined by
\BG
\label{b2}
\scalar{{\bf f}}{{\bf g}} := 
\frac{1}{4\pi}\int\limits_{0}^{\pi}d\vartheta\sin\vartheta\int\limits_{0}^{2\pi}d\varphi\,
\rom{f}(\rho,\vartheta,\varphi)^{*}\rom{g}(\rho,\vartheta,\varphi) \, ,
\EG 
where $\rom{f},\rom{g}$ are representants of ${\bf f}$ and ${\bf g}$, expressed in spherical
coordinates. The normalized (highest weight) spherical harmonics
\BG
\label{b3}
\Sh{j}{j} := \frac{\sqrt{\Gamma(2j+2)}}{2^{j}\Gamma(j+1)\rho^{j}}\lp
{\bf x}^{1}+i{\bf x}^{2}\rp^{j} \, , \qquad j\in\NNN,
\EG
are highest weight vectors of $\cdot^{(\rom{S})}$ with highest weight $j$, the restriction
of $\cdot^{(\rom{S})}$ to the $\Sl{2}$-submodules 
\BG
\label{b4}
{\bf V}^{j} := U(\Sl{2})^{(\rom{S})}\Sh{j}{j} 
\EG
is irreducible and the dense graded subalgebra $\Poloc{\Sp}\subseteq\Cuoc{\Sp}$ can be decomposed as $\Sl{2}$-module according to
\BG
\label{b5}
\Poloc{\Sp} = \bigoplus_{j\in\NNNa}{\bf V}^{j} \, .
\EG
The (ordinary) spherical harmonics, given by
\BG
\label{b6}
\Sh{j}{m} := \sqrt{\frac{\Gamma(j+m+1)}{\Gamma(2j+1)\Gamma(j-m+1)}}
\rom{J}_{-}^{(\rom{S})\, j-m}\Sh{j}{j} \, ,
\EG
where $m\in\{-j,-j+1,\cdots,j\}$, are orthonormal and they form a basis of ${\bf V}^{j}$
for every fixed $j\in\NNN$.\newline
As in the case of the supersphere one establishes on each truncated direct sum
\BG
\label{b7}
{\cal C}_{q} := \bigoplus_{j\in\NNNa \atop j\leq q}{\bf V}^{j} \, , \qquad q\in\NNN,
\EG
of $\Sl{2}$-modules a (noncommutative) associative product. In order to do so, let us
denote by $\cdot^{(\frac{q}{2})}:\Sl{2}\longrightarrow\End{\rom{V}(\frac{q}{2})}$ the irreducible $\Sl{2}$-representation with highest weight $\frac{q}{2}$, 
by $\{\rom{e}^{(\frac{q}{2})}_{m}\}_{m\in\{-\frac{q}{2},\cdots,\frac{q}{2}\}}$ the canonical
basis of $\rom{V}(\frac{q}{2})$ corresponding with the weight space decomposition and by $\scalar{\cdot}{\cdot}$ the scalar product on $\rom{V}(\frac{q}{2})$ fixed by the
requirement, that $\{\rom{e}^{(\frac{q}{2})}_{m}\}_{m\in\{-\frac{q}{2},\cdots,\frac{q}{2}\}}$ becomes an orthonormal basis. Then 
\BG
\label{b8}
\scalar{\rom{f}}{\rom{g}}_{\frac{q}{2}} := 
\frac{1}{q+1}\rom{Tr}\lp\rom{f}^{\dagger}\rom{g}\rp =
\frac{1}{q+1}\sum_{m=-\frac{q}{2}}^{m=\frac{q}{2}}\scalar{\rom{e}^{(\frac{q}{2})}_{m}}
{\rom{f}^{\dagger}\rom{g}\,\rom{e}^{(\frac{q}{2})}_{m}}
\EG
is a scalar product on $\End{\rom{V}(\frac{q}{2})}$. Moreover, denoting by 
$\DER{\End{\rom{V}(\frac{q}{2})}}$ the \CC-Lie algebra of derivations 
of the \CC-algebra $\End{\rom{V}(\frac{q}{2})}$, the representation 
$\rom{ad}^{(\frac{q}{2})}:\Sl{2}\longrightarrow\DER{\End{\rom{V}(\frac{q}{2})}}$, defined via
\BG
\label{b9}
\rom{ad}^{(\frac{q}{2})}\rom{J}(\rom{f}) := \Comu{\rom{J}^{(\frac{q}{2})}}{\rom{f}}
\EG
for all $\rom{J}\in\Sl{2}, \rom{f}\in\End{\rom{V}(\frac{q}{2})}$, becomes a star representation with respect to (\ref{b8}).\newline
For all $q\in\NNN$ and all $j\in\NNN, j\leq q$, the normalized endomorphisms
\BG
\label{b10}
\NSh{\frac{q}{2}}{j}{j} := \sqrt{\frac{2^{j}(2j+1)!!(q+1)\Gamma(q-j+1)}
{\Gamma(j+1)\Gamma(q+j+2)}}{\bf J}_{+}^{(\frac{q}{2})\, j}
\EG
are highest weight vectors of the representations $\rom{ad}^{(\frac{q}{2})}$, the
corresponding $\Sl{2}$-submodules
\BG
\label{b11}
\rom{V}^{(\frac{q}{2})\, j} := 
\rom{ad}^{(\frac{q}{2})}\lp U(\Sl{2})\rp\NSh{\frac{q}{2}}{j}{j}
\EG
are irreducible and $\End{\rom{V}(\frac{q}{2})}$ can be decomposed as $\Sl{2}$-module
according to
\BG
\label{b12}
\End{\rom{V}(\frac{q}{2})} = 
\bigoplus_{j\in\NNs_{0} \atop j\leq q}\rom{V}^{(\frac{q}{2})\, j} \, .
\EG
For every $q\in\NN$ we can introduce noncommutative spherical harmonics by
\BG
\label{b13}
\NSh{\frac{q}{2}}{j}{m} :=
\sqrt{\frac{\Gamma(j+m+1)}{\Gamma(2j+1)\Gamma(j-m+1)}}
\lp\rom{ad}^{(\frac{q}{2})}\rom{J}_{-}\rp^{j-m}\NSh{\frac{q}{2}}{j}{j} \, ,
\EG
where $j\in\NNN, j\leq q$ and $m\in\{-j,\cdots,j\}$, and they form an
orthonormal basis of $\End{\rom{V}(\frac{q}{2})}$.\newline
Now we can define isometric isomorphisms 
$\hat{\psi}_{q}:{\cal C}_{q}\longrightarrow\End{\rom{V}(\frac{q}{2})}$ of
$\Sl{2}$-modules by
\BG
\label{b14}
\hat{\psi}_{q}(\Sh{j}{m}) := \NSh{\frac{q}{2}}{j}{m} \, , \qquad
j\in\NNN, j\leq q, m\in\{-j,\cdots,j\} \, ,
\EG
as well as injective $\Sl{2}$-module homomorphism
$\hat{\eta}_{q'q}:\End{\rom{V}(\frac{q}{2})}\longrightarrow\End{\rom{V}(\frac{q'}{2})},
q\leq q',$ via
\BG
\label{b15}
\hat{\eta}_{q'q} := \hat{\psi}_{q'}\circ\hat{\iota}_{q'q}\circ\hat{\psi}_{q}^{-1} \, ,
\EG
where $\hat{\iota}_{q'q}$ denote the canonical injections 
${\cal C}_{q}\longrightarrow{\cal C}_{q'}$. Then $({\cal C}_{q},\hat{\iota}_{q'q})$ and
$(\End{\rom{V}(\frac{q}{2})},\hat{\eta}_{q'q})$ are isomorphic directed systems of
$\Sl{2}$-modules and their direct limits can be identified with $\Poloc{\Sp}$. 
Moreover the directed system $(\End{\rom{V}(\frac{q}{2})},\hat{\eta}_{q'q})$ possesses
a commutative limit: That is, for two arbitrary ${\bf f},{\bf f}'\in\Poloc{\Sp}$ the 
sequence $\{({\bf ff}')_{q}\}_{p\leq q,q\in\NNa}$ with
\BG
\label{b16}
({\bf ff}')_{q} := \hat{\eta}_{q}\lp\hat{\psi}_{q}({\bf f})\hat{\psi}_{q}({\bf f}')\rp =
\hat{\psi}_{q}^{-1}\lp\hat{\psi}_{q}({\bf f})\hat{\psi}_{q}({\bf f}')\rp \, ,
\EG
where $\hat{\eta}_{q}$ are the ``limit homomorphisms'' $\End{\rom{V}(\frac{q}{2})}\longrightarrow\Poloc{\Sp}$ and $p$ is some number, such that
${\bf f},{\bf f}'\in{\cal C}_{p}$, converges to ${\bf ff}'\in\Poloc{\Sp}$ with respect to the Fr\'{e}chet topology of $\Cuoc{\Sp}$.\newline
The directed system $(\End{\rom{V}(\frac{q}{2})},\hat{\eta}_{q'q})$ is called (2-dimensional)
fuzzy sphere; each one of the \CC-algebras and $\Sl{2}$-modules $\End{\rom{V}(\frac{q}{2})}$
is called truncated (2-dimensional) sphere and we denote it by $\NSp{q}$. Similar to the
truncated supersphere one finds for all $q\in\NN$
\BG
\label{b17}
\hat{\psi}_{q}({\bf x}^{k}) = 
\frac{2\rho}{\sqrt{q(q+2)}}\rom{J}^{(\frac{q}{2})}_{k} =: 
\rom{X}^{(\frac{q}{2})}_{k} \, , \qquad k=1,2,3,
\EG
and
\BG
\label{b18}
\sum_{k=1}^{3}\hat{\psi}_{q}({\bf x}^{k})^{2} =
\sum_{k=1}^{3}\lp\rom{X}^{(\frac{q}{2})}_{k}\rp^{2} = 
\rho^{2}\rom{Id}_{V(\frac{q}{2})} \, .
\EG

In order to introduce a natural noncommutative analogue to the body map 
$\beta_{\Ss}:\Huoc{\Ss}$\newline$\longrightarrow\Cuoc{\Sp}$ we note, that $\beta_{\Ss}$ is a 
$\Sl{2}$-module homomorphism because of (\ref{s16b}) and (\ref{s16c}). Moreover the restriction of $\beta_{\Ss}$ to $\Poloc{\Ss}$ is (respectively induces) a
surjective homomorphism
\BG
\label{b19}
\beta_{\Ss}\big\vert_{\Poloc{\Ss}}:\Poloc{\Ss}\longrightarrow\Poloc{\Sp}
\EG
of $\Sl{2}$-modules and \ZZg-graded \CC-algebras and we have explicitly
\BG
\label{b20}
\beta_{\Ss}(\Sphard{j}{l}{m}{2j}{\mu}) = \lbr
\begin{array}{ll}
\DS\frac{(-1)^{\mu}}{\sqrt{2l+1}}\Sh{l}{m} \, , & \quad \Over{2j+\mu}=\Over{0} ,
\\
0 \, , & \quad \Over{2j+\mu}=\Over{1} ,
\end{array}
\right.
\EG
for all $m\in\{-l,\cdots,l\}$. Consequently
\BG
\label{b21}
\beta_{\NSs{q}} := \hat{\psi}_{q}\circ\beta_{\Ss}\circ\psi_{q}^{-1}:
\NSs{q}\longrightarrow\NSp{q}
\EG
is a well-defined surjective $\Sl{2}$-module homomorphism for all $q\in\NN$ and we will
call it (noncommutative) body map of the truncated ($(2\vert 2)$-dimensional) supersphere.
Beside the nice feature
\BGG
\label{b22}
\beta_{\NSs{q}}(X^{(\frac{q}{2})}_{k}) &=& \rom{X}^{(\frac{q}{2})}_{k} \, , \qquad
k=1,2,3,
\nonumber
\\
\beta_{\NSs{q}}(\Theta^{(\frac{q}{2})}_{\alpha}) &=& 0 \, , \qquad
\alpha=4,5,
\EGG
there are some other immediate consequences, which suggest this interpretation.
\begin{proposition}
\label{theo1b}
$(\beta_{\NSs{q}})$ is a surjective homomorphism 
$(\NSs{q},\eta_{q'q})\longrightarrow(\NSp{q},\hat{\eta}_{q'q})$ of directed systems and 
$\lim\limits_{\longrightarrow}\beta_{\NSs{q}}$ is simply given by 
$\beta_{\Ss}\vert_{\Poloc{\Ss}}$. Moreover 
\BG
\label{b23}
\lim_{q\longrightarrow\infty}\hat{\eta}_{q}\lp\beta_{\NSs{q}}\lp
\psi_{q}(\Quotgr{f})\psi_{q}(\Quotgr{f}')\rp\rp =
\lim_{q\longrightarrow\infty}\lp\beta_{\Ss}(\Quotgr{f})\beta_{\Ss}(\Quotgr{f}')\rp_{q}
\EG
is fulfilled for all $\Quotgr{f},\Quotgr{f}'\in\Poloc{\Ss}$.
\end{proposition}
${\sl Proof}:$ For all $q,q'\in\NN, q\leq q'$ we have 
$\beta_{\NSs{q'}}\circ\eta_{q'q}=\hat{\eta}_{q'q}\circ\beta_{\NSs{q}}$ as well as
$\beta_{\Ss}\vert_{\Poloc{\Ss}}\circ\eta_{q}=\hat{\eta}_{q}\circ\beta_{\NSs{q}}$ by construction, which proves the first part. The ``homomorphism property in the limit''
follows according to
\BA
\lim_{q\longrightarrow\infty}\hat{\eta}_{q}\lp\beta_{\NSs{q}}\lp
\psi_{q}(\Quotgr{f})\psi_{q}(\Quotgr{f}')\rp\rp =
\lim_{q\longrightarrow\infty}\beta_{\Ss}\lp(\Quotgr{f}\Quotgr{f}')_{q}\rp =
\beta_{\Ss}(\Quotgr{f}\Quotgr{f}') =
\\
= \beta_{\Ss}(\Quotgr{f})\beta_{\Ss}(\Quotgr{f}') =
\lim_{q\longrightarrow\infty}\lp\beta_{\Ss}(\Quotgr{f})\beta_{\Ss}(\Quotgr{f}')\rp_{q}
\qquad \qquad 
\EA
where we used the first part of the proposition, the (graded-)commutative limit of the fuzzy (super)sphere and the continuity of $\beta_{\Ss}$.
\hfill $\Box$ 

\section{A graded differential calculus on the fuzzy supersphere}

Let $X$ be a \Hu-deWitt supermanifold.
We introduced the elements of the \CC-Lie superalgebra and \ZZg-graded $\Huoc{X}$-module $\Svec{X}$ of complex, global supervector fields on $X$ as graded derivations of
the \ZZg-graded \CC-algebra $\Huoc{X}$. The latter concept can be generalized to
arbitrary \ZZg-graded \CC-algebras without any change, when we replace ``\ZZg-graded $\Huoc{X}$-module'' by the formulation ``\ZZg-graded module over the graded center $\Cent{\Huoc{X}}$ of the \ZZg-graded \CC-algebra $\Huoc{X}$'', which is equivalent in the 
graded-commutative case. 
Moreover, again viewing $X$ as graded manifold, \NNN-homogeneous, complex, global superdifferential forms are by definition graded-alternating $\Huoc{X}$-multilinear, or
equivalently $\Cent{\Huoc{\Ss}}$-multilinear maps from $\Svec{X}$ to the \ZZg-graded \CC-algebra of \Hu-functions $\Huoc{X}$ \cite{Bartocci1,Kostant1}. Adopting the
graded center-formulation we can generalize the notion of superdifferential forms to
arbitrary \ZZg-graded \CC-algebras and take it as starting point for the development
of a graded differential calculus. This is the basic idea of the construction of 
derivation-based differential calculi 
\cite{Dubois-Violette1,Dubois-Violette3,Dubois-Violette5,Dubois-Violette6}, transferred to the 
\ZZg-graded case.\newline
This idea can even be generalized by taking into account only a subset of all 
graded derivations, which is at the same time a \CC-Lie subsuperalgebra as well as a graded submodule over the graded center of the \ZZg-graded \CC-algebra under consideration. 
We will employ, as it was done in the ungraded ``fuzzy case'' 
\cite{Madore2,Madore10,Grosse7}, a specific variant of this generalization to develop an analogue
of the super-deRham complex and the graded Cartan calculus on each of the truncated superspheres. A very natural feature of this approach will be, that the noncommutative
body map extends - as in the case of graded manifolds - to a cochain map from the 
differential complex on the truncated supersphere to the one on its body.
 
According to the general principles formulated above the first thing we need to know is
the graded center $\Cent{\NSs{q}}$ of the truncated superspheres $\NSs{q}, q\in\NN$.
Analogous to the ungraded case (see for example \cite{Pierce1}) we find, that 
$\NSs{q}$ is graded-central,
\BG 
\label{d1}
\Cent{\NSs{q}} = \Centg{\NSs{q}}{0} = \CC\rom{Id}_{V(\frac{q}{2},\Over{1})} \, . 
\EG
Consequently one can choose (according to the argumentation above) in principle every \CC-Lie subsuperalgebra of $\Derc{\NSs{q}}$ as \CC-Lie superalgebra of supervector fields $\Svec{\NSs{q}}$ on each truncated supersphere. But there is a natural choice given by the action of the orthosymplectic Lie superalgebra $\Ort{1}{2}$ on $\NSs{q}$, 
\BG
\label{d2}
\Svec{\NSs{q}} := \rom{ad}^{(\frac{q}{2})}\lp\Ort{1}{2}\rp \, , \quad q\in\NN.
\EG
The graded representations $\rom{ad}^{(\frac{q}{2})}$ are faithful by (\ref{t12}),
such that there are natural isomorphisms
\BG
\label{d3}
\Svec{\NSs{q}} \cong \Svec{\NSs{q'}} \cong \Ort{1}{2} \, , \quad q,q'\in\NN,
\EG
of \CC-Lie superalgebras and this fact will ``control the growth of the graded
differential calculus with respect to $q$''.\newline 
An additional justification of the choice (\ref{d2}) stems from a translation of the
$\Sl{2}$-module homomorphism property of $\beta_{\NSs{q}}$:
We can define a map $\tilde{\beta}_{\NSs{q}}$ from $\Svecg{\NSs{q}}{0}$ to the 
\CC-Lie algebra $\Vec{\NSp{q}}:=\rom{ad}^{(\frac{q}{2})}(\Sl{2})\cong\Sl{2}$ of (complex) vector fields on the truncated sphere $\NSp{q}$ \cite{Madore2,Madore10} via
\BG
\label{d3a}
\tilde{\beta}_{\NSs{q}}(D)\beta_{\NSs{q}}(f) := \beta_{\NSs{q}}(Df) \, , \qquad
f\in\NSs{q},
\EG
analogous to (\ref{s16c}). Then
$\tilde{\beta}_{\NSs{q}}$ is a surjective (in fact a bijective) Lie algebra homomorphism, which maps $\rom{ad}^{(\frac{q}{2})}J\in\Svecg{\NSs{q}}{0}$ to
\BG
\label{d3b}
\tilde{\beta}_{\NSs{q}}(\rom{ad}^{(\frac{q}{2})}J) = \rom{ad}^{(\frac{q}{2})}\rom{J} \, ,
\EG
where $\rom{J}\in\Sl{2}$ corresponds to $J\in\Ortg{1}{2}{0}$ via the natural 
isomorphism.

For every natural number $p\in\NN$ let us denote by $\Hrpmm{\CCa}{p}{\Svec{\NSs{q}}}{\NSs{q}}$
the \ZZg-graded \CC-vector space of all $p$-linear maps
$\Svec{\NSs{q}}\times\stackrel{_p}{\ldots}\times\Svec{\NSs{q}}\longrightarrow\NSs{q}$ and by $\Perm{p}$ the symmetric group on $p$ letters.
Introducing the commutation factor 
$\gamma_{p}:\Perm{p}\times\ZZg\times\stackrel{_p}{\ldots}\times\ZZg\longrightarrow\{\pm 1\}$
via
\BG
\label{d4}
\gamma_{p}(\sigma;\Over{i}_{1},\cdots,\Over{i}_{p}) :=
\prod_{r,s=1,\cdots,p;r<s\atop
\sigma^{-1}(r)>\sigma^{-1}(s)}
(-1)^{\Over{i}_{r}\Over{i}_{s}} \, ,
\EG
we can define a representation $\pi$ of $\Perm{p}$ on $\Hrpmm{\CCa}{p}{\Svec{\NSs{q}}}{\NSs{q}}$ by
\BG
\label{d5}
\lp\pi_{\sigma}\omega\rp(D_{1},\cdots,D_{p}) :=
\gamma_{p}(\sigma;\Over{D}_{1},\cdots,\Over{D}_{p})
\omega(D_{\sigma(1)},\cdots,D_{\sigma(p)})
\EG
for all $\omega\in\Hrpmm{\CCa}{p}{\Svec{\NSs{q}}}{\NSs{q}},\sigma\in\Perm{p}$ and all homogeneous $D_{1},\cdots,D_{p}\in\Svec{\NSs{q}}$ \cite{Scheunert2}. 
Now by definition a $p$-linear map $\omega\in\Hrpmm{\CCa}{p}{\Svec{\NSs{q}}}{\NSs{q}}$ is called graded-alternating if
\BG
\label{d6}
\pi_{\sigma}\omega = \rom{sgn}\sigma\,\omega
\EG
is fulfilled for all $\sigma\in\Perm{p}$ and, according to the discussion at the beginning
of the section, we should interpret them as $p$-superforms on the truncated
supersphere $\NSs{q}$. The set of all $p$-superforms on the truncated
supersphere $\NSs{q}$ forms a graded vector subspace of $\Hrpmm{\CCa}{p}{\Svec{\NSs{q}}}{\NSs{q}}$ and it will be denoted by $\SDCp{p}{q}$.\newline
A general superform on the truncated supersphere $\NSs{q}$ is an element of the direct sum
\BG
\label{d7}
\SDC{q} := \bigoplus_{p\in\NNNa}\SDCp{p}{q} \, ,
\EG 
where we set $\SDCp{0}{q}:=\NSs{q}$. Employing the multiplicative structure of $\NSs{q}$
we can proceed exactly as in the case of graded manifolds 
\cite{Bartocci1,Kostant1} (respectively graded Lie-Cartan pairs \cite{Jadczyk1,Jadczyk2,Pittner1}) to introduce a graded wedge product on $\SDC{q}$. So we define first
for all $p,p'\in\NNN,\Over{i},\Over{i'}\in\ZZg$ a bilinear map
$\wedge:\SDCpg{p}{q}{i}\times\Omega^{g,p'}({\cal S}_{\tiny\rho,q})_{\Over{i'}}\longrightarrow
\Omega^{g,p+p'}({\cal S}_{\tiny\rho,q})_{\Over{i}+\Over{i'}}$ by
\BGG
\label{d8}
\lp\omega\wedge\omega'\rp(D_{1},\cdots,D_{p+p'}) :=
\frac{1}{p!p'!}\sum_{\sigma\in\Perma{p+p'}}\rom{sgn}\sigma\,
\gamma_{p+p'}(\sigma;\Over{D}_{1},\cdots,\Over{D}_{p+p'})\cdot
\qquad \qquad 
\\
\cdot(-1)^{\Over{i'}\sum_{l=1}^{p}\Over{D}_{\sigma(l)}}
\omega(D_{\sigma(1)},\cdots,D_{\sigma(p)})
\omega'(D_{\sigma(p+1)},\cdots,D_{\sigma(p+p')})
\nonumber
\EGG
for all homogeneous $D_{1},\cdots,D_{p+p'}\in\Svec{\NSs{q}}$ and extend these by
bilinearity to $\SDC{q}$. With respect to it $\SDC{q}$ becomes a $\NNN\times\ZZg$-bigraded
\CC-algebra.

Having built up the algebra of superforms on each truncated supersphere we can introduce the graded Cartan calculus as one does it for ordinary graded manifolds 
\cite{Kostant1,Bartocci1}. As far as we are only interested in the linear structure of $\SDC{q}$ we are doing nothing else than Lie superalgebra cohomology of $\Svec{\NSs{q}}$ with values in $\NSs{q}$ \cite{Fuks1,Scheunert8}. So let us follow the excellent
article \cite{Scheunert8} and apply it to our case.\newline
One first extends the action of the \CC-Lie superalgebra $\Svec{\NSs{q}}$ to $\SDC{q}$:
For every homogeneous $D\in\Svec{\NSs{q}}$ one introduces a \CC-linear map
$L_{D}:\SDC{q}\longrightarrow\SDC{q}$ by defining its action on bihomogeneous
$\omega\in\SDCp{p}{q}, p\in\NNN,$ according to
\BGG
\label{d9}
\lp L_{D}\omega\rp(D_{1},\cdots,D_{p}) &:=&
D\lp\omega(D_{1},\cdots,D_{p})\rp -
\\
& & -\sum_{l=1}^{p}(-1)^{\Over{D}(\Over{\omega}+\sum_{l'=1}^{l-1}\Over{D}_{l'})} 
\omega(D_{1},\cdots,\Comug{D}{D_{l}},\cdots,D_{p})
\nonumber
\EGG
for all homogeneous $D_{1},\cdots,D_{p}\in\Svec{\NSs{q}}$. $L_{D}$ is a bihomogeneous endomorphism of the bigraded \CC-vector space $\SDC{q}$ of bidegree $(0,\Over{D})$. A general graded derivation $D\in\Svec{\NSs{q}}$ can be uniquely decomposed
into its homogeneous components $D_{i},i=0,1,$ and 
$L_{D}:=L_{D_{0}}+L_{D_{1}}\in\End{\SDC{q}}$ is well-defined. Moreover, the map
$L:\Svec{\NSs{q}}\longrightarrow\End{\SDC{q}}$, 
given by $D\mapsto L_{D}$, is a graded representation of the
\CC-Lie superalgebra $\Svec{\NSs{q}}$ on $\SDC{q}$.\newline
Besides this extension of $\Svec{\NSs{q}}$ it is useful to introduce for all
$D\in\Svec{\NSs{q}}$ a \CC-linear map $\imath_{D}:\SDC{q}\longrightarrow\SDC{q}$ by
\BG
\label{d10}
\imath_{D}f := 0
\EG
for all $f\in\NSs{q}$ and by
\BG
\label{d11}
\lp\imath_{D}\omega\rp(D_{2},\cdots,D_{p}) := \omega(D,D_{2},\cdots,D_{p})
\EG
for all $\omega\in\SDCp{p}{q}, p\in\NN$ and $D_{2},\cdots,D_{p}\in\Svec{\NSs{q}}$.
For every $D\in\Svec{\NSs{q}}$ it is a \NNN-homogeneous of degree $-1$ and the map
$\imath:\Svec{\NSs{q}}\longrightarrow\End{\SDC{q}},D\mapsto\imath_{D}$ is \CC-linear and
\ZZg-even. In addition the relations
\BG
\label{d12}
\imath_{D}\circ\imath_{D'} + (-1)^{\Over{D}\Over{D'}}\imath_{D'}\circ\imath_{D} = 0
\EG
and
\BG
\label{d13}
\lp L_{D}\circ\imath_{D'}-\imath_{D'}\circ L_{D}\rp\omega =
(-1)^{\Over{D}\Over{\omega}}\imath_{\Comug{D}{D'}}\omega
\EG
are fulfilled for all homogeneous $D,D'\in\Svec{\NSs{q}}$ and all \ZZg-homogeneous
$\omega\in\SDC{q}$.\newline
The Lie superalgebra cochain map is most elegantly introduced as \CC-linear map
$d:\SDC{q}\longrightarrow\SDC{q}$, whose action on \ZZg-homogeneous $0$-superforms
$f\in\NSs{q}$ is given by 
\BG
\label{d14}
df(D) := (-1)^{\Over{f}\Over{D}}Df
\EG
for all homogeneous $D\in\Svec{\NSs{q}}$ and whose action on \ZZg-homogeneous $p$-superforms
$\omega\in\SDCp{p}{q},p\in\NN,$ is defined inductively via
\BG
\label{d15}
\imath_{D}\lp d\omega\rp := (-1)^{\Over{D}\Over{\omega}}L_{D}\omega - d\lp\imath_{D}\omega\rp
\EG
for all homogeneous $D\in\Svec{\NSs{q}}$. Then $d$ is a bihomogeneous endomorphism of 
bidegree $(1,\Over{0})$ fulfilling the cochain condition
\BG
\label{d16}
d\circ d = 0
\EG
and commuting with $L_{D}$ for all $D\in\Svec{\NSs{q}}$. Explicitly one finds
\BGG
\label{d17}
d\omega(D_{0},\cdots,D_{p}) =
\sum_{l=0}^{p}
(-1)^{l+\Over{D}_{l}(\Over{\omega}+\sum_{l'=0}^{l-1}\Over{D}_{l'})}
L_{D_{l}}\lp\omega(D_{0},\cdots,\stackrel{\vee}{D_{l}},\cdots,D_{p})\rp + \qquad
\\
+\sum_{0\leq l<l'\leq p}
(-1)^{l'+\Over{D}_{l'}\sum_{l''=l+1}^{l'-1}\Over{D}_{l''}}
\omega(D_{0},\cdots,D_{l-1},\Comug{D_{l}}{D_{l'}},\cdots,\stackrel
{\vee}{D_{l'}},\cdots,D_{p})
\nonumber
\EGG
for homogeneous $D_{0},D_{1},\cdots,D_{p}\in\Svec{\NSs{q}}$
and bihomogeneous $\omega\in\SDC{q}$ of bidegree $(p,\Over{\omega})$
($\vee$ denotes omission).\newline
The preceeding discussion can be summarized as follows: As far as one considers only the
linear structure of $\SDC{q}$ the endomorphisms $L_{D},\imath_{D}$ and $d$ fulfill exactly
the same relations as Lie derivative, inner product and exterior derivative in the case
of graded manifolds. The latter observation stays also true if one considers the graded
wedge product (\ref{d8}) on $\SDC{q}$.
\begin{proposition} 
\label{theo1d}
The relations
\BGG
\label{d18}
L_{D}\lp\omega\wedge\omega'\rp &=&
\lp L_{D}\omega\rp\wedge\omega' + (-1)^{\Over{D}\Over{\omega}}\omega\wedge L_{D}\omega'
\nonumber
\\
\imath_{D}\lp\omega\wedge\omega'\rp &=&
(-1)^{\Over{D}\Over{\omega'}}\lp\imath_{D}\omega\rp\wedge\omega' + 
(-1)^{p}\omega\wedge\imath_{D}\omega'
\\
d\lp\omega\wedge\omega'\rp &=&
\lp d\omega\rp\wedge\omega' + (-1)^{p}\omega\wedge d\omega'
\nonumber
\EGG
are fulfilled for all homogeneous $D\in\Svec{\NSs{q}}$ and all bihomogeneous
$\omega,\omega'\in\SDC{q}$ of bidegree $(p,\Over{\omega})$ and $(p',\Over{\omega'})$,
respectively. 
\end{proposition}
${\sl Proof}$: This can be shown exactly as in the case of graded manifolds. That is, one
starts with a direct proof of the second relation and proofs the other equations
inductively using the relations (\ref{d13}) and (\ref{d15}).
\hfill $\Box$ 

\vspace{2mm}

Because of the analogy to the case of graded manifolds (see \cite{Kostant1}) we will call the endomorphisms $L_{D}, \imath_{D}$ and $d$ of $\SDC{q}$ Lie derivative and interior product (with respect to the supervector field $D\in\Svec{\NSs{q}}$) as well as exterior derivative.

With respect to the product $\wedge$ the \ZZg-graded \CC-vector space $\SDCp{p}{q},p\in\NNN$,
of $p$-superforms forms a \ZZg-graded $\NSs{q}$-bimodule. In the special case 
$\SDCp{0}{q}\equiv\NSs{q}$ this \ZZg-graded (left as well as right) module is graded-free
with homogeneous basis $\{\rom{Id}_{V(\frac{q}{2},\Over{1})}\}$, of course.
In order to investigate the other \ZZg-graded $\NSs{q}$-bimodules $\SDCp{p}{q},p\in\NN$,
a little bit closer, let 
$\{~E_{k}\in\Ortg{1}{2}{0},E_{\alpha}\in\Ortg{1}{2}{1}~\vert~k=1,2,3;\alpha=4,5~\}$ be 
some homogeneous basis of $\Ort{1}{2}$ and 
\BG
\label{d19}
\partial_{q,A} := \rom{ad}^{(\frac{q}{2})}E_{A} \, , \qquad A=1,\cdots,5,
\EG
the elements of the corresponding homogeneous basis of $\Svec{\NSs{q}}$. Denoting by
$\zeta^{A}_{q}$ the elements of the dual basis to $\{\partial_{q,A}\}$ we can
introduce homogeneous 1-superforms $\lambda^{A}_{q}\in\SDCp{1}{q},A=1,\cdots,5,$ via
\BG
\label{d20}
\lambda^{A}_{q}(D) := \zeta^{A}_{q}(D)\rom{Id}_{V(\frac{q}{2},\Over{1})}
\EG
for all $D\in\Svec{\NSs{q}}$. By applying both sides on supervector fields we find
\BGG
\label{d21}
\lambda^{A}_{q}\wedge\lambda^{B}_{q} &=&
-(-1)^{\Over{\lambda}^{A}_{q}\Over{\lambda}^{B}_{q}}\lambda^{B}_{q}\wedge\lambda^{A}_{q}
\nonumber
\\
f\wedge\lambda^{A_{1}}_{q}\wedge\cdots\wedge\lambda^{A_{p}}_{q} &=&
(-1)^{\Over{f}\sum_{l=1}^{p}\Over{\lambda}^{A_{l}}_{q}}
\lambda^{A_{1}}_{q}\wedge\cdots\wedge\lambda^{A_{p}}_{q}\wedge f 
\EGG
for all homogeneous $f\in\NSs{q}$ and all $A,B,A_{1},\cdots,A_{p}=1,\cdots,5, p\in\NN$, as
well as
\BG
\label{d22}
d\lambda^{A}_{q} = 
\frac{1}{2}\sum_{B,C=1}^{5}c_{BC}^{A}\lambda^{C}_{q}\wedge\lambda^{B}_{q} \, ,
\EG
where $c_{BC}^{A}$ denote the $\Ort{1}{2}$-structure constants corresponding with the
basis $\{E_{A}\}$. Introducing the graded vector subspace $\SDCpc{p}{q}, p\in\NN$, of all
$p$-superforms with values in the graded center $\Cent{\NSs{q}}$ of $\NSs{q}$, we can
conclude from (\ref{d21}) (or from the usual isomorphisms between graded-alternating maps and
the graded exterior algebra \cite{Bartocci1,Scheunert2,Scheunert7,Scheunert8}), that
\BG
\label{d23}
\lbr~\lambda^{A_{1}}_{q}\wedge\cdots\wedge\lambda^{A_{p}}_{q}~\big
\vert~(A_{1},\cdots,A_{p})\in\GIndn{p}\rbr
\EG
with
\BGG
\label{d24}
\GIndn{p} := 
\lbr~(A_{1},\cdots,A_{p'},A_{p'+1},\cdots,A_{p})~\big\vert~0\leq p'\leq p;
A_{1},\cdots,A_{p'}=1,2,3; 
\right.
\qquad \qquad \qquad
\\
\left.
A_{p'+1},\cdots,A_{p}=4,5; 
A_{1}<A_{2}<\cdots<A_{p'}<A_{p'+1}\leq\cdots\leq A_{p-1}\leq A_{p}~\rbr
\nonumber
\EGG
forms a homogeneous basis of $\SDCpc{p}{q}$. Moreover the bigraded subalgebra of $\SDCp{p}{q}$,
\BG
\label{d25}
\SDCc{q} := \bigoplus_{p\in\NNNa}\SDCpc{p}{q} 
\EG
with $\SDCpc{0}{q}:=\Cent{\NSs{q}}$, is stable under the Lie derivative, the interior
product as well as under exterior differentiation according to (\ref{d22}).\newline
From the preceding discussion we can conclude in particular, that the \ZZg-graded
$\NSs{q}$-modules $\SDCp{p}{q}, p\in\NN,$ are graded-free (as left and as right modules) and
a homogeneous basis is given by (\ref{d23}). Consequently every $\omega\in\SDCp{p}{q}$
can be written as
\BG
\label{d26}
\omega = \sum_{(A_{1},\cdots,A_{p})\in\GIndn{p}}
\omega_{A_{1}\cdots A_{p}}\wedge\lambda^{A_{1}}_{q}\wedge\cdots\wedge\lambda^{A_{p}}_{q}
\EG
and the unique coefficients $\omega_{A_{1}\cdots A_{p}}\in\NSs{q}$ are given by
\BG
\label{d27}
\omega_{A_{1}\cdots A_{p}} = 
(-1)^{\frac{1}{2}p''(p''-1)}\frac{1}{\prod_{A=1}^{5}N_{A}!}
\omega(\partial_{q,A_{1}},\cdots,\partial_{q,A_{p}}) \, ,
\EG
where $p''$ is the number of entries in $(A_{1},\cdots,A_{p})$ greater than 3 and $N_{A}$ 
is the number of entries in $(A_{1},\cdots,A_{p})$ being equal $A$.

Similar to the case of matrix geometry (\cite{Dubois-Violette1,Madore10}) it is possible to introduce via
\BG
\label{d28}
\Lambda_{q} := \sum_{A=1}^{5}E^{(\frac{q}{2})}_{A}\wedge\lambda_{q}^{A}
\EG 
an even 1-superform on each truncated supersphere $\NSs{q}, q\in\NN$, which is invariant 
and fulfills a super-version of the Maurer-Cartan equation.
\begin{proposition}
\label{theo2d}
The definition of $\Lambda_{q}$ is independent of the choice of the homogeneous basis of
$\Ort{1}{2}$. $\Lambda_{q}$ is invariant and up to complex multiples it is the only invariant
1-superform on $\NSs{q}, q\in\NN$. Moreover, its exterior differential fulfills
\BG
\label{d29}
d\Lambda_{q} = \Lambda_{q}\wedge\Lambda_{q}
\EG
and the exterior differential of each $f\in\SDCp{0}{q}$ can be written according to
\BG
\label{d30}
df = \Comug{\Lambda_{q}}{f} \equiv \Lambda_{q}\wedge f - f\wedge\Lambda_{q} \, .
\EG
\end{proposition}
${\sl Proof}:$ Beside the uniqueness statement only simple calculations are involved.
In order to see, that $\Lambda_{q}$ is the only invariant 1-superform on $\NSs{q}$ it
is important to note, that $\SDCpc{1}{q}$ is an irreducible, \ZZg-graded 
$\Ort{1}{2}$-module with highest weight 1 and that the \ZZg-graded 
$\Ort{1}{2}$-module $\SDCp{1}{q}$ is isomorphic to the tensor product of the
\ZZg-graded $\Ort{1}{2}$-modules $\NSs{q}$ and $\SDCpc{1}{q}$. Then the uniqueness of
$\Lambda_{q}$ follows from proposition \ref{theo1t} and the ``Clebsch-Gordan decomposition''
of tensor products of (irreducible) \ZZg-graded $\Ort{1}{2}$-modules \cite{Berezin2}.
\hfill $\Box$ 

\vspace{2mm}

If $X$ is a \Hu-deWitt supermanifold with body X the body map $\beta_{X}$ extends to
an algebra homomorphism and cochain map from the super-deRham complex of $X$ to the
ordinary deRham complex on the body manifold X \cite{Kostant1}. Because the noncommutative
body map $\beta_{\NSs{q}}:\NSs{q}\longrightarrow\NSp{q}, q\in\NN$, is no algebra
homomorphism, some extension from the algebra of superforms on the truncated
supersphere to the algebra of forms on the truncated sphere cannot be an algebra 
homomorphism. But by translating the construction of \Hu-deWitt supermanifolds (respectively
graded manifolds) we will introduce a cochain map, which we can interpret as
noncommutative analogue to the extension of $\beta_{X}$ in the graded-commutative 
setting.

Let us give first of all definition and basic results of the Cartan calculus (see
\cite{Dubois-Violette1,Madore2,Madore10}) on the truncated sphere $\NSp{q}, q\in\NN$, as far as they are relevant for the subsequent discussion of the body map of superforms on the truncated supersphere $\NSs{q}$.\newline
For every $p\in\NN$ a $p$-form on the truncated sphere $\NSp{q}$ is a $p$-linear, alternating
map $\Vec{\NSp{q}}\times\stackrel{_p}{\ldots}\times\Vec{\NSp{q}}\longrightarrow\NSp{q}$ 
and we denote by $\DCp{p}{q}$ the \CC-vector space of all $p$-forms. A general form
on the truncated sphere $\NSp{q}$ is an element of the direct sum
\BG
\label{d31}
\DC{q} := \bigoplus_{p\in\NNNa}\DCp{p}{q}
\EG
with $\DCp{0}{q}:=\NSp{q}$. $\DC{q}$ becomes a \NNN-graded \CC-algebra, if
one introduces for all $p,p'\in\NNN$ a bilinear map
$\hwedge:\DCp{p}{q}\times\DCp{p'}{q}\longrightarrow\DCp{p+p'}{q}$ by
\BG
\label{d32}
\lp\hat{\omega}\hwedge\hat{\omega}'\rp(\rom{D}_{1},\cdots,\rom{D}_{p+p'}) :=
\frac{1}{p!p'!}\sum_{\sigma\in\Perma{p+p'}}\rom{sgn}\sigma\,
\hat{\omega}(\rom{D}_{\sigma(1)},\cdots,\rom{D}_{\sigma(p)})
\hat{\omega}'(\rom{D}_{\sigma(p+1)},\cdots,\rom{D}_{\sigma(p+p')})
\EG
for all $\rom{D}_{1},\cdots,\rom{D}_{p+p'}\in\Vec{\NSp{q}}$, and extends these by
bilinearity. Analogous to the graded case the set $\DCpc{p}{q}, p\in\NN$, of $p$-forms with values in the center $\Centc{\NSp{q}}=\CC\rom{Id}_{\rom{V}(\frac{q}{2})}$ of $\NSp{q}$ forms
a vector subspace of $\DCp{p}{q}$ and $\DCc{q}:=\oplus_{p\in\NNNa}\DCpc{p}{q},
\DCpc{0}{q}:=\Centc{\NSp{q}}$, is a graded subalgebra of $\DC{q}$.\newline
Exterior differential d, Lie derivative $\rom{L}_{\rom{D}}$ as well as interior product
$\i_{\rom{D}}$ (with respect to a vector field $\rom{D}\in\Vec{\NSp{q}}$) are defined
exactly as in the case of the truncated supersphere, if one views $\Vec{\NSp{q}}$ and
$\DC{q}$ as trivially \ZZg-graded.\newline
For some basis $\{ \rom{E}_{k} \vert k=1,2,3 \}$ of $\Sl{2}$ one can introduce a basis
$\{ \hat{\partial}_{q,k} \vert k=1,2,3 \}$ of $\Vec{\NSp{q}}$ as well as 1-forms
$\hat{\lambda}^{k}_{q}\in\DCpc{1}{q}$ analogous to (\ref{d19}) and (\ref{d20}). 
The latter 1-forms anticommute and one finds, that the set of $p$-forms
\BG
\label{d33}
\lbr~\hat{\lambda}^{k_{1}}_{q}\hwedge\cdots\hwedge\hat{\lambda}^{k_{p}}_{q}~\big
\vert (k_{1},\cdots,k_{p})\in\OIndn{p}~\rbr
\EG
with
\BG
\label{d34}
\OIndn{p} := \lbr~(k_{1},\cdots,k_{p})~\big
\vert~k_{1},\cdots,k_{p}=1,2,3; k_{1}<\cdots<k_{p}~\rbr
\EG
forms for all $p\in\NN$ (trivially for $p>3$, of course) a basis of the (left and right) $\NSp{q}$-module $\DCp{p}{q}$ as well as of the \CC-vector space $\DCpc{p}{q}$.

In analogy to the case of graded manifolds \cite{Kostant1} we can introduce now maps $\beta^{(p)}_{\NSs{q}}:\SDCp{p}{q}\longrightarrow\DCp{p}{q}, p\in\NNN,$
via
\BG
\label{d35}
\lp\beta^{(p)}_{\NSs{q}}(\omega)\rp\lp
\tilde{\beta}_{\NSs{q}}(D_{1}),\cdots,\tilde{\beta}_{\NSs{q}}(D_{p})\rp :=
\beta_{\NSs{q}}\lp\omega\lp D_{1},\cdots,D_{p}\rp\rp 
\EG
for all $D_{1},\cdots,D_{p}\in\Svecg{\NSs{q}}{0}$. In the case $p=0$ (\ref{d35}) is to be 
understood as $\beta^{(0)}_{\NSs{q}}=\beta_{\NSs{q}}$, of course. For each $p\in\NNN$ it 
is an even \CC-linear map and we can define its \CC-linear extension 
$\SDC{q}\longrightarrow\DC{q}$, which we again denote by $\beta_{\NSs{q}}$.
\begin{proposition}
\label{theo3d}
$\beta_{\NSs{q}}$ is a surjective $\Sl{2}$-module homomorphism and cochain map,
whose restriction to $\SDCc{q}$ is an algebra homomorphism onto $\DCc{q}$.
\end{proposition}
${\sl Proof}:$ Using the $\Sl{2}$-module homomorphism property of the body map and
the fact, that $\tilde{\beta}_{\NSs{q}}$ is a Lie algebra isomorphism one
finds immediately
\BGG
\label{d36}
&& \rom{d}\circ\beta_{\NSs{q}} = \beta_{\NSs{q}}\circ d \qquad \qquad
\nonumber
\\
&& \rom{L}_{\tilde{\beta}_{\NSs{q}}(D)}\circ\beta_{\NSs{q}} = 
\beta_{\NSs{q}}\circ L_{D} \qquad \qquad
\EGG
for all $D\in\Svecg{\NSs{q}}{0}$.
The action of the extension of the body map on a $p$-superform $\omega\in\SDCp{p}{q}$
can be described alternatively by
\BGG
\label{d37}
\beta_{\NSs{q}}(\omega) \equiv 
\beta_{\NSs{q}}\lp\sum_{(A_{1},\cdots,A_{p})\in\GIndna{p}}
\omega_{A_{1}\cdots A_{p}}\wedge\lambda^{A_{1}}_{q}\wedge\cdots\wedge\lambda^{A_{p}}_{q}\rp =
\nonumber
\\
= \sum_{(A_{1},\cdots,A_{p})\in\OIndna{p}}
\beta_{\NSs{q}}\lp\omega_{A_{1}\cdots A_{p}}\rp\hwedge\hat
{\lambda}^{A_{1}}_{q}\hwedge\cdots\hwedge\hat{\lambda}^{A_{p}}_{q} \, ,
\qquad \qquad
\EGG
where $\lambda^{k}_{q}\in\SDCp{1}{q}$ and $\hat{\lambda}^{k}_{q}\in\DCp{1}{q}$ correspond
to bases in $\Ortg{1}{2}{0}$ and $\Sl{2}$, which are related by the canonical isomorphism.
From (\ref{d36}) we can conclude in particular, that $\beta_{\NSs{q}}$ is surjective and
that its restriction to $\SDCc{q}$ is an algebra homomorphism.
\hfill $\Box$ 

\vspace{2mm}

The interpretation of the \ZZg-graded \CC-algebras and \ZZg-graded $\Ort{1}{2}$-modules
$\NSs{q}$ stems from the fact, that there are suitable $\Ort{1}{2}$-module homomorphisms
$\eta_{q'q}:\NSs{q}\longrightarrow\NSs{q'}, q,q'\in\NN, q\leq q'$, constituting a
directed system, whose direct limit can be identified with the graded subalgebra
$\Poloc{\Ss}\subseteq\Huoc{\Ss}$ and which possesses a graded-commutative limit.
We want to understand noncommutative $p$-superforms for $p\in\NN$ in a similar way as 
noncommutative pendants to $p$-superforms on the $(2\vert 2)$-dimensional supersphere
$\Ss$. Although we reserve the precise treatment of this question to a subsequent paper
we give some indications how this is done.\newline
The first thing one has to find are maps 
$\eta_{q'q}^{(p)}:\SDCp{p}{q}\longrightarrow\SDCp{p}{q'}, q,q'\in\NN, q\leq q'$,
such that $(\SDCp{p}{q},\eta_{q'q}^{(p)}), p\in\NNN,$ become directed systems, which 
are ``compatible with the Cartan calculus''. 
A natural choice for these maps (especially in consideration of a ``graded-commutative 
limit'' of superforms) is 
\BGG
\label{d38}
\eta_{q'q}^{(p)}(\omega) \equiv 
\eta_{q'q}^{(p)}\lp\sum_{(A_{1},\cdots,A_{p})\in\GIndna{p}}
\omega_{A_{1}\cdots A_{p}}\wedge\lambda^{A_{1}}_{q}\wedge\cdots\wedge\lambda^{A_{p}}_{q}\rp :=
\nonumber
\\
:= \sum_{(A_{1},\cdots,A_{p})\in\GIndna{p}}
\eta_{q'q}\lp\omega_{A_{1}\cdots A_{p}}\rp\wedge
\lambda^{A_{1}}_{q'}\wedge\cdots\wedge\lambda^{A_{p}}_{q'} \, .
\qquad \qquad
\EGG
Then $(\SDCp{p}{q},\eta_{q'q}^{(p)}), p\in\NNN,$ are directed systems of \ZZg-graded
$\Ort{1}{2}$-modules, whose direct limits should be connected with the algebra of superforms
on the $(2\vert 2)$-dimensional supersphere. This problem is in fact in exactly the same way
``singular'' as in the case of the ordinary fuzzy sphere \cite{Madore2,Madore10}.
It is more natural to interpret the elements of the direct limits as superforms on the
supergroup UOSP$(1\vert 2)$ (see \cite{Bartocci10,Berezin2}, for example), which is
a superfibre bundle over the $(2\vert 2)$-dimensional supersphere.

\section{Cohomological considerations}

According to a general theorem of \cite{Kostant1} the super-deRham cohomology of the
$(2\vert 2)$-dimensional supersphere is isomorphic to the deRham cohomology of the 
$2$-dimensional sphere. The isomorphism is induced by extension of the body map (in the
sense described in the preceding chapter) to the algebra of superforms on the 
$(2\vert 2)$-dimensional supersphere. As we will see exactly the same is true on
every truncated supersphere.

The cohomology 
\BG
\label{c1}
H(\NSs{q}) \equiv \bigoplus_{p\in\NNNa}H^{p}(\NSs{q}) := \frac{\rom{ker}d}{\rom{im}d}
\EG
of the complex $(\SDC{q},d)$ is our substitute for the (complexified) super-deRham cohomology on the truncated supersphere $\NSs{q}, q\in\NN$, as well as the cohomology
\BG
\label{c2}
H(\NSp{q}) \equiv \bigoplus_{p\in\NNNa}H^{p}(\NSp{q}):= \frac{\rom{ker}\,\rom{d}}{\rom{im}\,\rom{d}}
\EG
of the complex $(\DC{q},\rom{d})$ is the noncommutative pendant to the (complexified) deRham cohomology of the sphere. By construction $H(\NSs{q})$ is the Lie superalgebra cohomology of
$\Svec{\NSs{q}}$ with coefficients in $\NSs{q}$, while $H(\NSp{q})$ is the Lie algebra cohomology of $\Vec{\NSp{q}}$ with coefficients in $\NSp{q}$.\newline
The cochain map $\beta_{\NSs{q}}$ induces as usual via
\BGG
\label{c3}
H(\beta_{\NSs{q}}): H(\NSs{q})\longrightarrow H(\NSp{q})
\nonumber
\\
\lb\omega\rb\mapsto\lb\beta_{\NSs{q}}(\omega)\rb \quad \;\;
\EGG
a homomorphism of \NNN-graded \CC-vector spaces, which turns out to be an isomorphism.
\begin{proposition}
\label{theo1c}
$H(\beta_{\NSs{q}})$ is an isomorphism of bigraded \CC-vector spaces and we have
explicitly
\BG
\label{c4}
H^{p}(\NSs{q}) \cong H^{p}(\NSp{q}) \cong \lbr
\begin{array}{ll}
\CC \, , & \qquad p=0,3 \\
\{ 0\} \, , & \qquad p\in\NNN\setminus\{ 0,3\} \, .
\end{array} 
\right.
\EG
\end{proposition}
${\sl Proof}:$ Because the (graded) representations $\rom{ad}^{(\frac{q}{2})}$ are faithful,
$H(\NSs{q})$ is isomorphic to the Lie superalgebra cohomology $H(\Ort{1}{2};\NSs{q})$ of
$\Ort{1}{2}$ with coefficients in $\NSs{q}$ as well as $H(\NSp{q})$ is isomorphic to the Lie algebra cohomology $H(\Sl{2};\NSp{q})$ of $\Sl{2}$ with coefficients in $\NSp{q}$.
Moreover, using the direct sum decompositions (\ref{t12}) and (\ref{b12}), we find
\cite{Scheunert8}
\BD
H^{p}(\NSs{q}) \cong 
\bigoplus_{j\in\frac{1}{2}\NNNs\atop j\leq q}H^{p}(\Ort{1}{2};\NVSphar{\frac{q}{2}}{j})
\ED
and
\BD
H^{p}(\NSp{q}) \cong
\bigoplus_{j\in\NNs_{0} \atop j\leq q}H^{p}(\Sl{2};\rom{V}^{(\frac{q}{2})\, j}) \, .
\ED
The standard second-order Casimir operator of $\Ort{1}{2}$, restricted to
$\NVSphar{\frac{q}{2}}{j},j\neq 0,$ has a nonvanishing eigenvalue and there are no
nontrivial $\Ort{1}{2}$-invariant elements in $\NVSphar{\frac{q}{2}}{j},j\neq 0,$
such that we can conclude \cite{Scheunert8}
\BD
H^{p}(\NSs{q}) \cong 
H^{p}(\Ort{1}{2};\NVSphar{\frac{q}{2}}{0}) \cong
H^{p}(\Ort{1}{2};\CC) \, ,
\ED
where $H(\Ort{1}{2};\CC)$ denotes the Lie superalgebra cohomology of $\Ort{1}{2}$ with
trivial coefficients. The same argument leads in the case of the truncated sphere to
\BD
H^{p}(\NSp{q}) \cong 
H^{p}(\Sl{2};\CC) \, .
\ED
But these cohomologies with trivial coefficients are known to be given by (\ref{c4})
(see \cite{Fuks1,Fuks2} for the case of $\Ort{1}{2}$ and \cite{Greub1} for
$\Sl{2}$).  
\hfill $\Box$ 

\vspace{2mm}

One should note, that (\ref{c4}) is not exactly the super-deRham cohomology (deRham cohomology) of the $(2\vert 2)$-dimensional supersphere (2-dimensional sphere). The
explicit result is rather a cohomological verification of the remarks we made at the 
end of the preceding chapter about the singular character of the identification of the
noncommutative ``limit superforms'' with superforms on the $(2\vert 2)$-dimensional
supersphere.

\section{Concluding Remarks}

In close analogy to the construction of the fuzzy sphere and its derivation-based
differential calculus we have introduced the fuzzy supersphere together with a
differential calculus, which is based on the Lie superalgebra $\Ort{1}{2}$
acting on each of the truncated superspheres via graded derivations. The natural interpretation of the fuzzy sphere as noncommutative body of the fuzzy supersphere guaranteed the existence 
of the usual relations between \Hu-deWitt supermanifolds and their bodies. In particular
the noncommutative body projection induced in the same way as in the theory of graded 
manifolds the isomorphism between the ((graded) derivation-based) cohomologies of the 
fuzzy supersphere and its body.\newline
The present work can be seen as first step towards the development of the differential
geometry and the formulation of (quantum) field theoretical models on a non-trivial,
(fuzzy) noncommutative supermanifold. From the point of view of pure supergeometry the
introduction of metric and supervector bundle concepts as well as the investigation of
their ``limits'' are the next challenging tasks. 
Passing from mathematics to (quantum) physical model building we will have to undertake a
``categorial jump'' from \ZZg-graded algebras of \Hu-functions and their noncommutative
approximations to larger classes of ``functions''. In our opinion the most promising
program in that direction is a G-extension, graded-commutative and noncommutative as
well.
 
\section*{Acknowledgement}

The authors would like to thank W.Bulla, C.Klim\v{c}ik, H.Miglbauer, F.Pauer, P.Pre\v{s}najder, L.Pittner, A.Strohmaier and W.Tutschke for helpful discussions and the
``Fonds zur F\"orderung der wissenschaftlichen Forschung (FWF)'' for 
support funding.

\end{document}